\documentclass[a4paper,fleqn]{cas-dc}

\usepackage[authoryear]{natbib}
\usepackage{subcaption}   
\usepackage[percent]{overpic} 
\setkeys{Gin}{draft=false}
\usepackage{bm}

\def\tsc#1{\csdef{#1}{\textsc{\lowercase{#1}}\xspace}}
\tsc{WGM}
\tsc{QE}
\tsc{EP}
\tsc{PMS}
\tsc{BEC}
\tsc{DE}

\makeatletter
\ExplSyntaxOn

\tex_gdef:D \abstractname { \textbf{Abstract} }

\RenewDocumentEnvironment{Abstract}{o}
 {
   \group_begin:
   \IfNoValueTF{#1}{}{ \tex_gdef:D \abstractname{#1} }
   \parindent\z@
   \noindent \rule{\textwidth}{.2pt}\par
   \vspace{6pt}
   \noindent \abstractname \par
   \vspace{6pt} 
   \ignorespaces
 }
 { \par \vspace{8pt} \group_end: } 

\RenewDocumentEnvironment{keywords}{O{Keywords}}
 {
   \cs_new:Nn \sep: { ;\c_space_token }
   \cs_set_eq:NN \sep \sep:
   \tex_global:D \tex_setbox:D \g_stm_key_box = \vtop \bgroup
     \hsize=\textwidth \parindent\z@
     \noindent \textit{#1\keywordtitlesep}
 }
 { \par \egroup }

\RenewDocumentCommand \MaketitleBox { O{} }
 {
  \tl_if_blank:nTF { #1 }{}{ \keys_set:nn { stm / mktitle } { #1 } }
  \processbreakafter
  \tex_def:D \baselinestretch{1}
  \group_begin: \@title \group_end:
  \bool_if:NTF \g_stm_blind_bool
    { \vspace*{10mm} }
    {
      \group_begin:
      \normalsize \stmauthors \par
      \stmcollab \par
      \footnotesize \itshape \stmaddress \par
      \group_end:
      \bool_if:NTF \g_stm_breakafter_auaff_bool { \newpage }{}
    }
  \begin{Abstract}
    \noindent \ignorespaces
    \file_if_exist:nT {\jobname.abs} { \file_input:n {\jobname.abs} }
  \end{Abstract}
  \noindent \box \g_stm_key_box
  \par\vspace{6pt}
  \noindent \rule{\textwidth}{.2pt}\par
  \vspace{12pt} 
  \bool_if:NTF \g_stm_breakafter_abstract_bool { \newpage }{}
 }

\RenewDocumentCommand \LongMaketitleBox { O{} }
 {
  \tl_if_blank:nTF { #1 }{}{ \keys_set:nn { stm / mktitle } { #1 } }
  \vbox_gset:Nn \g_stm_front_box
   {
     \tex_def:D \baselinestretch{1}
     \group_begin: \@title \group_end:
     \bool_if:NTF \g_stm_blind_bool
       { \vspace*{10mm} }
       { \group_begin:
         \normalsize \stmauthors \par
         \stmcollab \par
         \footnotesize \itshape \stmaddress \par
         \group_end: }
     \begin{Abstract}
       \noindent \ignorespaces
       \file_if_exist:nT {\jobname.abs} { \file_input:n {\jobname.abs} }
     \end{Abstract}
     \noindent \box \g_stm_key_box
     \par\vspace{6pt}
     \noindent \rule{\textwidth}{.2pt}\par
     \vspace{12pt}
   }
  \vbox_gset:Nn \g_stm_notes_box
   { \cs_set_eq:NN \footnotetext \__fn_text:n \printFirstPageNotes }
  \dim_gset:Nn \g_tmpb_dim { \box_ht:N \g_stm_notes_box }
  \dim_gadd:Nn \g_tmpb_dim { \box_dp:N \g_stm_notes_box }
  \ifbool{sc}{\dim_gadd:Nn \g_tmpb_dim { 12pt }}{}
 }
\ExplSyntaxOff
\makeatother

\makeatletter
\ExplSyntaxOn
\cs_set:Npn \__first_footerline:
  {
    \group_begin:
    \small
    \sffamily
    \ifnum\theblind>0\relax
    \else 
      \__short_authors: :~
    \fi
    { \rmfamily \itshape Preprint}
    \group_end:
  }
\ExplSyntaxOff
\makeatother

\begin{document}
\let\WriteBookmarks\relax
\def\floatpagepagefraction{1}
\def\textpagefraction{.001}
\shorttitle{FractMorph: A Fractional Fourier-Based Multi-Domain Transformer for Deformable Image Registration}
\shortauthors{S. Kebriti et~al.}

\title [mode = title]{FractMorph: A Fractional Fourier-Based Multi-Domain Transformer for Deformable Image Registration}

\author[1]{Shayan Kebriti}[orcid=0009-0008-4446-0806]%

\author[1]{Shahabedin Nabavi}[orcid=0000-0001-7240-0239]%
\cormark[1]

\author[2,3]{Ali Gooya}[orcid=0000-0001-5135-4800]%

\affiliation[1]{%
  organization={Faculty of Computer Science and Engineering, Shahid Beheshti University},%
  city={Tehran},%
  country={Iran}%
}

\affiliation[2]{%
  organization={School of Computing Science, University of Glasgow},%
  city={Glasgow},%
  country={UK}%
}

\affiliation[3]{%
  organization={Alan Turing Institute},%
  city={London},%
  country={UK}%
}

\cortext[1]{Corresponding author. \emph{E-mail address:} s\_nabavi@sbu.ac.ir}

\begin{abstract}
Deformable image registration (DIR) is a crucial and challenging technique for aligning anatomical structures in medical images and is widely applied in diverse clinical applications. However, existing approaches often struggle to capture fine-grained local deformations and large-scale global deformations simultaneously within a unified framework.
We present FractMorph, a novel 3D dual-parallel transformer-based architecture that enhances cross-image feature matching through multi-domain fractional Fourier transform (FrFT) branches. Each Fractional Cross-Attention (FCA) block applies parallel FrFTs at fractional angles of $0^\circ$, $45^\circ$, $90^\circ$, along with a log-magnitude branch, to effectively extract local, semi-global, and global features at the same time. These features are fused via cross-attention between the fixed and moving image streams. A lightweight U-Net style network then predicts a dense deformation field from the transformer-enriched features.
On the intra-patient ACDC cardiac MRI dataset, FractMorph achieves state-of-the-art performance with an overall Dice Similarity Coefficient (DSC) of $86.45\%$, an average per-structure DSC of $75.15\%$, and a 95th-percentile Hausdorff distance (HD95) of $1.54~\mathrm{mm}$ on our data split. FractMorph-Light, a lightweight variant of our model with only 29.6M parameters, preserves high accuracy while halving model complexity.
Furthermore, we demonstrate the generality of our approach with solid performance on a cerebral atlas-to-patient dataset.
Our results demonstrate that multi-domain spectral–spatial attention in transformers can robustly and efficiently model complex non-rigid deformations in medical images using a single end-to-end network, without the need for scenario-specific tuning or hierarchical multi-scale networks. The source code is available at \href{https://github.com/shayankebriti/FractMorph}{https://github.com/shayankebriti/FractMorph}.
\end{abstract}

\begin{keywords}
Deep learning \sep Vision Transformer \sep Image registration \sep Medical image analysis \sep Fractional Fourier transform
\end{keywords}

\maketitle

\section{Introduction}
\label{sec:introduction}
Image registration is the process of estimating spatial transformations to align the corresponding anatomical structures of multiple images \citep{ZITOVA2003977}. This technique is widely used in clinical applications, since aligning images taken at various times, orientations, or modalities is needed for accurate medical analysis and diagnosis.
Registration methods are generally categorized according to the type of transformation they perform. Rigid registration is used for translation and rotation transformations, whereas affine registration extends this also to include scaling and shearing \citep{1242337}. A single 2D matrix is sufficient to represent these transformations. Although both rigid and affine approaches have been shown to be effective in practice, they become less accurate when the shape of anatomical structures changes significantly between images \citep{zou2022review}. Deformable image registration (DIR) overcomes this limitation by facilitating non-rigid deformations within the images \citep{1242337}. This flexibility allows DIR to model detailed changes in anatomical structures accurately. A deformable transformation is commonly represented as a dense deformation field, formulated as a 3D matrix in 2D image registration tasks and a 4D matrix for 3D cases \citep{zou2022review}. Correctly estimating these complex, high-dimensional matrices is a challenging task because of the complexity of capturing large and irregular shape changes in anatomical structures, especially when the deformation is massive or highly irregular.

Deformable registration has shown significant value in a wide range of clinical uses, including diagnosis, surgical planning, and longitudinal monitoring. It forms the foundation of numerous critical tasks, such as identification of tumor progression, organ tracking, surgical navigation, and multi-organ mapping \citep{ramadan2024medical}. It is especially valuable for integrating information from varied imaging modalities, such as combining the soft tissue contrast of MRI with the bone detail of CT, enabling a more comprehensive anatomical and functional understanding that enhances diagnostic accuracy and treatment planning \citep{huang2020review}.
DIR also plays a crucial role in analyzing anatomical structures by allowing detailed motion tracking of tissues. In cardiac imaging, accurate assessment of cardiac strain, derived from DIR, is more sensitive than traditional measures such as left ventricular ejection fraction (LVEF), especially in detecting early signs of cardiac failure, including heart failure with preserved ejection fraction (HFpEF) \citep{pfeffer2019heart}. Although commonly used in clinical practice, cine MRI sequences such as steady-state free precession (SSFP) do not directly encode tissue motion \citep{bistoquet2008myocardial}. DIR allows for the extraction of deformation-based metrics from cine SSFP MRI data, making advanced cardiac function analysis feasible without additional scans. These strain-based metrics are vital for the diagnosis of cardiomyopathies, the evaluation of valve diseases, the detection of regional dysfunctions, and the guidance of therapeutic interventions \citep{ARRATIALOPEZ2023102925}.

In this paper, we introduce a novel 3D transformer-based framework for deformable image registration. Unlike conventional CNNs, characterized by their emphasis on local receptive fields, transformers leverage the attention mechanism to learn long‑range correspondences within input features.
FractMorph processes fixed and moving volumes in parallel transformer streams, exchanging information via our Fractional Cross‑Attention (FCA) blocks. Each FCA block enriches feature maps through multi‑domain 3D fractional Fourier transform (FrFT) branches. These branches enable the model to capture spatial, mixed spatial-frequency, and frequency domains simultaneously. Finally, a lightweight U‑Net style network generates the dense deformation field while preserving detail and smoothness.
Compared to existing DIR methods, which struggle to identify both local and global characteristics efficiently, our proposed framework addresses this gap by understanding the local, semi-global, and global correspondences and deformations simultaneously. The key contributions of this study are listed below:
\begin{itemize}
    \item We propose FractMorph, a novel 3D Transformer-based framework that fuses multi-scale local, semi-global, and global features at every stage for accurate deformable image registration. A lightweight encoder-decoder CNN follows the Transformer to generate high-resolution, detailed deformation fields.
    \item We introduce the FCA module, which combines multi-order 3D FrFT branches with convolutional and attention mechanisms to enrich feature representations across spatial, spectral, and fractional domains.
    \item We develop an efficient separable implementation of 3D FrFT and integrate it into our framework, demonstrating improved registration performance with feasible computational cost. To the best of our knowledge, this is the first integration of 3D FrFT into deep learning models for medical imaging.
    \item We achieve registration accuracy that outperforms current state-of-the-art and recently published approaches using both our main and lightweight variants on an intra-patient cardiac MRI benchmark dataset. Additionally, we demonstrate the generality and solid performance of our approach on a cerebral atlas-to-patient dataset.
\end{itemize}

The structure of this paper is as outlined below. In Section \ref{sec:related-work}, an overview of related studies is provided. We present the proposed framework in Section~\ref{sec:method}. Section~\ref{sec:experiments} presents statistical outcomes alongside qualitative analyses. We discuss these findings in Section~\ref{sec:discussion}. Lastly, we present our conclusions and suggest directions for future studies in Section~\ref{sec:conclusion}.

\begin{figure*}[!t]
\centering
\centerline{\includegraphics[width=\textwidth]{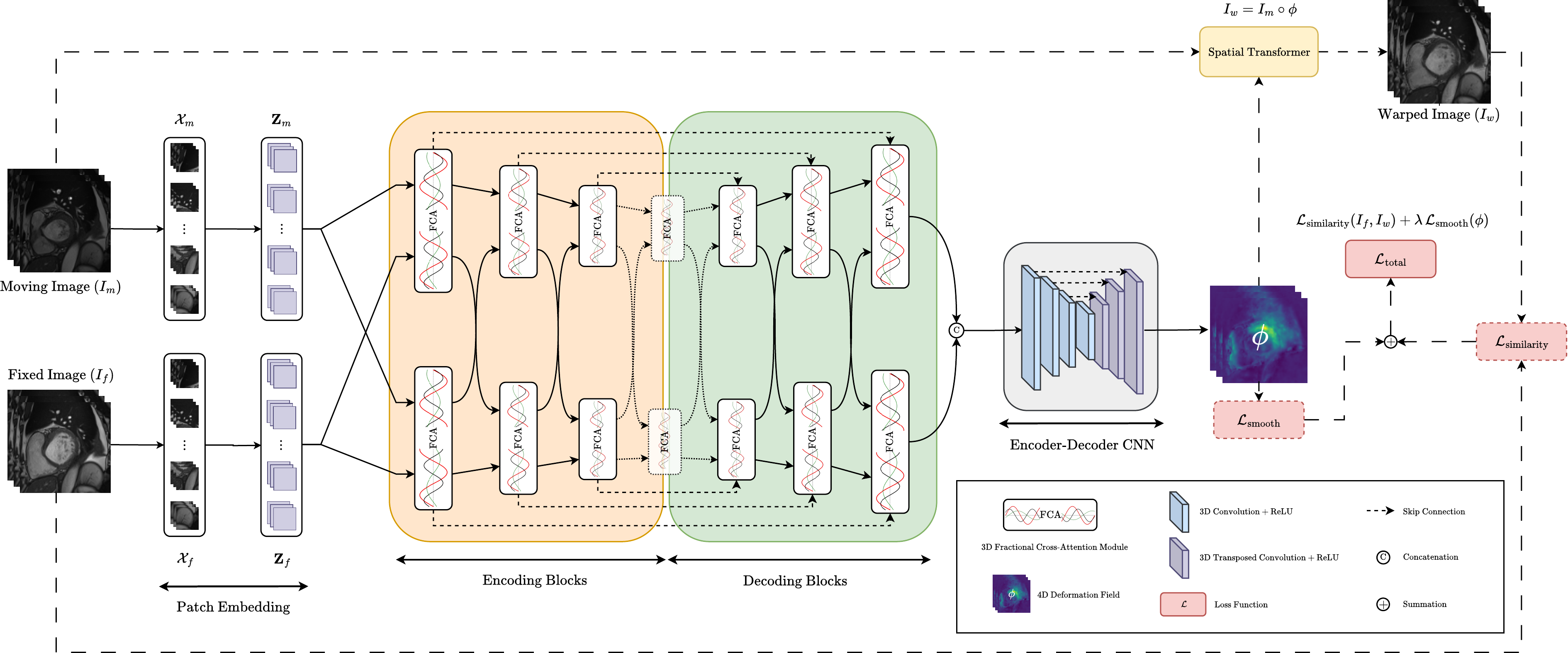}}
\caption{Overview of the proposed FractMorph framework. Given a fixed and a moving 3D image volume, both are first divided into non-overlapping patches and embedded as tokens. These token sequences are jointly processed by symmetric encoding and decoding transformer blocks. At each stage, FCA blocks enable bidirectional interaction between the fixed and moving streams, with skip connections linking encoding and decoding features. After decoding, the two streams are concatenated and passed to a lightweight 3D U-Net style network that outputs the displacement field $\phi$. The Spatial Transformer uses this field to warp the moving image toward the fixed image. Ultimately, a similarity loss is computed on the warped-fixed pair, combined with a regularization term for smoothness on the deformation field.}
\label{fig:fractmorph}
\end{figure*}

\section{Related Work}
\label{sec:related-work}
\subsection{Traditional Methods}
Traditional deformation-based alignment techniques have been extensively studied and widely applied in medical imaging over the past several decades. These techniques enable the alignment of complex anatomical structures across modalities and have become standard tools in applications such as neuroimaging, cardiac imaging, radiation therapy planning, and beyond.
Key examples include free-form deformation (FFD) algorithms utilizing B-spline models \citep{jiang2003fpga}, optical flow-based methods such as the Demons algorithm \citep{cahill2009demons}, and diffeomorphic approaches like Symmetric Normalization (SyN) \citep{avants2008symmetric}.
B-spline FFD methods model deformation as a continuous spline function constructed on a sparse control point mesh. Registration is typically performed by optimizing a cost function that combines an intensity-based similarity metric with a regularization term to suppress irregular transformations \citep{rueckert2002nonrigid}. B-spline FFDs are frequently employed in medical imaging applications due to their flexibility in modeling local deformations and robust performance across varying anatomies. However, achieving high registration accuracy often requires a fine control-point grid and a large number of optimization iterations, making the method computationally expensive for large 3D volumes.
The Demons algorithm represents a parameter-free approach to deformable image registration that draws an analogy to optical flow and diffusion processes. It iteratively updates a dense displacement field by computing force vectors based on intensity differences and applying a smoothing step at each iteration. In essence, the source image is gradually aligned with the target image by accumulating incremental displacements. While the Demons algorithm is efficient and effective for moderate deformations, it struggles with large anatomical differences.
Unlike traditional asymmetric approaches, the SyN algorithm performs symmetric optimization by computing halfway deformations from each image to a shared intermediate space, rather than warping one image entirely to the other. This formulation reduces bias toward either the fixed or moving image and yields inverse-consistent results. SyN has demonstrated excellent performance and has been widely adopted in applications such as neuroimaging, cardiac MRI, and abdominal CT. However, its primary drawback is high computational cost, as it requires substantial memory and runtime to execute effectively.
In practice, all these classical methods involve solving an iterative optimization for each new pair of images, and their performance can degrade or require manual tuning when faced with very large or highly localized deformations \citep{klein2009evaluation}.

\subsection{CNN-Based and Other Non-Transformer Methods}
Deep learning-based registration techniques have emerged in recent years to address limitations of traditional methods. These approaches accelerate the registration process by replacing per-case optimization with an end-to-end trainable network. A foundational contribution in this area is VoxelMorph \citep{balakrishnan2019voxelmorph}, which introduced an unsupervised convolutional U-Net architecture to predict dense displacement fields between a moving and fixed image. VoxelMorph achieved registration accuracy comparable to traditional methods while reducing computational cost. However, its accuracy still leaves room for improvement. To address this limitation, CycleMorph \citep{kim2021cyclemorph} introduced a cyclically consistent model designed to register images from A to B and vice versa simultaneously, encouraging deformations that preserve the topological structure.

Beyond convolutional architectures, researchers have explored alternative frameworks that introduce new approaches to modeling the deformation field. For instance, a recent work involves diffusion probabilistic modeling as an alternative to single-pass forward prediction. DiffuseReg \citep{zhuo2024diffusereg} employs a denoising diffusion model to iteratively learn the deformation field through diffusion steps. This approach improves interpretability as real-time observation of the registration process is possible. Furthermore, it allows users to inspect intermediate warped outputs and adjust the noise level, which offers more control over the registration behavior. Though DiffuseReg achieved higher accuracy than prior diffusion-based registration models, its inference time remains considerably slower than single-pass methods.
\citep{jia2023fourier} explores frequency-domain representations for more efficient registration by proposing Fourier-Net. They replace the U-Net architecture's expansive path with a parameter-free, model-driven decoder that operates in a band-limited Fourier domain. Also, rather than estimating a high-resolution displacement field within the spatial space, it learns a compact Fourier formulation. By adopting this approach, the design significantly reduces both the parameter count and the number of multiply-add operations.

Another example of alternative frameworks is WarpPINN \citep{ARRATIALOPEZ2023102925}, which incorporates physics-informed neural networks into DIR that formulates the task with biomechanical constraints. In WarpPINN, the network directly predicts the displacement field by enforcing the limited compressibility of heart tissue by penalizing the Jacobian determinant associated with the deformation. Additionally, the use of Fourier feature mapping of the input images helps WarpPINN overcome its spectral bias and better capture high-frequency deformations.
A major challenge here arises from large non-rigid deformations and the complex structure of certain anatomies, such as the heart. To address this, 
\cite{chang2024independently} presents a multi-scale registration model trained in isolation. In place of a singular end-to-end network, a hierarchical structure is implemented to sequentially capture large-scale deformations with increasing resolution. This framework consists of several separate V-Net architectures, each trained on different resolutions of the input images. At the inference stage, the moving image is initially processed by the low-resolution network. The predicted deformation field is then upsampled and passed through the next network, and so on. By decomposing the problem, this approach effectively addresses the challenge of very large motions. However, because each scale’s model is trained independently, there is no joint optimization across scales, which increases the risk of inconsistencies when fusing multi-scale deformations.
In general, although previous methods perform well in capturing local details, they often struggle to model global context effectively. So far, it remains challenging to find a straightforward approach to capture both local and global context simultaneously.

\begin{figure*}[!t]
\centering
\centerline{\includegraphics[width=\textwidth]{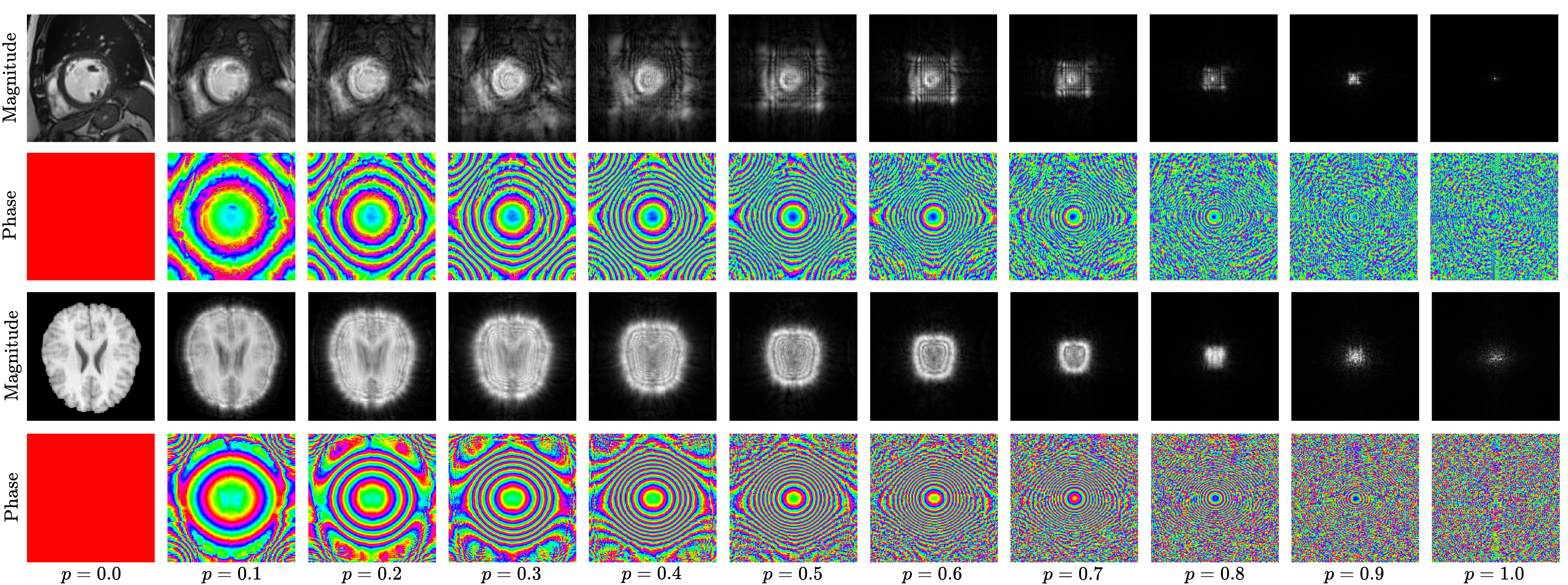}}
\caption{Magnitude and phase of our implemented 3D FrFT applied to cardiac (top two rows) and cerebral (bottom two rows) medical images at different fractional orders. For illustration, only a single 2D slice of each volume is shown.}
\label{fig:frft-progress}
\end{figure*}

\subsection{Transformer-Based Methods}
Transformers, as a new class of architectures, have recently gained significant attention in the research community. These models are distinguished by their proficiency in learning long-range and global dependencies through their attention mechanisms, enabling them to overcome the local receptive field limitations of previous network types.
An early example of Transformer usage in image registration is ViT-V-Net \citep{chen2021vit}, which integrates a Vision Transformer (ViT) encoder block into a V-Net architecture. This hybrid ConvNet–Transformer model achieved superior performance compared to VoxelMorph.
\cite{chen2022transmorph} later presented TransMorph, which leverages Swin Transformer-based encoders \citep{liu2021swin} to retrieve feature maps from the moving and fixed images, and a convolutional decoder to generate the corresponding displacement field. Notably, TransMorph was extended with diffeomorphic warping and a Bayesian variant for uncertainty estimation. TransMorph achieves higher alignment accuracy than previous pure CNN models and also outperforms ViT-V-Net.

More recent works have primarily focused on designing Transformers specialized for the DIR task. For example, Attention-Reg \citep{song2022cross} leverages a cross-modal attention mechanism that maps the features of one image to those of another, thereby overcoming the feature correspondence gap of previous methods and better capturing the mutual information between images.
Another example is XMorpher \citep{shi2022xmorpher}, which proposes a dual-parallel Transformer design that processes the moving and fixed images in parallel by keeping correspondence at each stage. It takes the moving and fixed images as separate inputs and learns their correspondence by sharing information via parallel cross-attention blocks. This approach achieves enhanced results over earlier methods, which either concatenated the images before passing them through a network or used separate networks for each image and fused the features afterward.
TransMatch \citep{chen2023transmatch} takes a different strategy to improve feature correspondence. It combines self-attention and cross-attention mechanisms in a two-stream architecture, where each image is first passed through separate self-attention blocks. The resulting features are then combined using cross-attention and passed to a decoder convolution architecture to generate the final deformation field. This approach enables the model to capture correspondences between the moving and fixed images in a hierarchical manner.
However, the field still lacks models that can fully and simultaneously address both local features and global context throughout the entire registration process. Moreover, there remain promising strategies that have not yet been explored in this area.

\section{Methodology}
\label{sec:method}
FractMorph is a deep 3D deformable image registration framework that incorporates a novel parallel transformer to fuse multi-domain features from the input images, followed by an encoder-decoder U-Net style architecture to generate a detailed deformation field. This framework takes a fixed image $I_f$ and a moving image $I_m$ as input and generates a deformation field $\phi$ that warps $I_m$ to align with $I_f$. Our proposed framework is designed to capture both local details and global contextual information by integrating convolutional processing, FrFT branches, and a cross-attention mechanism between $I_f$ and $I_m$. The overall pipeline of our work is illustrated in Fig.~\ref{fig:fractmorph}.

\subsection{Patch Embedding}
\label{subsec:patch_embedding}
The Patch Embedding module is the first component of the network and is responsible for converting the input 3D image volumes into structured sequences of feature tokens. It converts the input volumes fixed image (\( I_f \in \mathbb{R}^{H \times W \times D}\)) and moving image (\( I_m \in \mathbb{R}^{H \times W \times D}\)) into a set of feature map tokens which are suitable for transformer processing. The 3D volume of each of the two input images ($I_f$ and $I_m$) is partitioned into non-overlapping cubic blocks of dimensions $P_x \times P_y \times P_z$. 
This results in a total of $N = \frac{H \cdot W \cdot D}{P_x \cdot P_y \cdot P_z}$ patches per image.
Let us denote the set of 3D patches extracted from an image $I$ as:
\begin{equation}
\mathcal{X} = \{\,\mathbf{x}^1,\mathbf{x}^2,\ldots,\mathbf{x}^N\},\quad
\mathbf{x}^i\in\mathbb{R}^{P_x\times P_y\times P_z}
\end{equation}
Each patch $\mathbf{x}^i$ is then mapped to a $d$-dimensional token using a learnable linear projection matrix $\mathbf{E} \in \mathbb{R}^{P_x \times P_y \times P_z \times d}$. Specifically, each token is computed as:
\begin{equation}
\mathbf{z}^i = \mathbf{x}^i \mathbf{E}, \quad \mathbf{z}^i \in \mathbb{R}^d
\end{equation}
Embedded token sequences are obtained by applying this to all patches:
\begin{equation}
\mathbf{Z}_f = \{ \mathbf{z}_f^1, \mathbf{z}_f^2, \ldots, \mathbf{z}_f^N \}, \quad \mathbf{Z}_m = \{ \mathbf{z}_m^1, \mathbf{z}_m^2, \ldots, \mathbf{z}_m^N \}
\end{equation}
Here, $\mathbf{Z}_f$ and $\mathbf{Z}_m$ are the token representations of the fixed and moving volumes, respectively.
This patch-based embedding provides two main benefits. First, it reduces the spatial resolution of the input, enabling more efficient processing and attention across the volume. Second, by mapping each patch to a higher-dimensional and semantically more meaningful feature space, the network gains more expressive capacity for modeling structures.

\begin{figure}[!t]
\centerline{\includegraphics[width=\columnwidth]{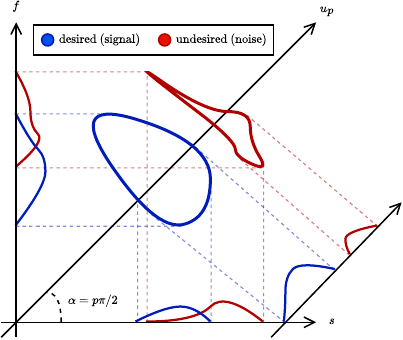}}
\caption{
FrFT-based separation of desired from undesired components in the spatial–spectral domain, where the spatial ($s$), frequency ($f$), and fractional order ($u_p$) axes are shown.
}
\label{fig:spatial-spectral-space}
\end{figure}

\subsection{Dual-Parallel Transformer}
\label{subsec:dual-transformer}
The core of our framework is a dual-input Transformer that processes the moving and fixed image tokens in parallel and enables explicit cross-feature matching between them. This architecture integrates our novel FCA mechanism for feature matching. Our transformer utilizes a dual-stream feature extraction approach, with one stream dedicated to the moving image and the other to the fixed image. In contrast to conventional backbone designs that either process a single image or use a single stream to handle the concatenation of two images, this dual-stream approach enables the network to exchange information between the moving and fixed images at multiple levels of representation. By propagating mutual information across multiple resolutions, the network can uncover multi-level semantic correspondences while simultaneously extracting features. This design facilitates more effective alignment, as features from $I_f$ and $I_m$ are progressively brought into correspondence at different scales.
Let $\mathbf{F}^l_m$ and $\mathbf{F}^l_f$ $(\mathbf{F}^l_{m}, \mathbf{F}^l_{f} \in \mathbb{R}^{H_l \times W_l \times D_l \times C_l})$ denote the moving and fixed feature maps at level $l$ of the encoder. Level $0$ takes the output of the patch embedding layer as its input.
The encoder of our transformer consists of $L$ levels with increasing channel dimensions and decreasing spatial resolution. At each level $l$, a series of Transformer blocks is applied. $n_l$ denotes the number of blocks at level $l$. Each Transformer block takes $(\mathbf{F}^l_m, \mathbf{F}^l_f)$ as input and and produces refined feature maps $(\widetilde{\mathbf{F}}^l_m,\, \widetilde{\mathbf{F}}^l_f)$ through dual cross-attention fusion. After processing each level $l$, the feature maps are downsampled by a scaling ratio of 2 across all spatial dimensions to produce the inputs $\mathbf{F}^{l+1}_m$ and $\mathbf{F}^{l+1}_f$ for the next level:
\begin{equation}
H_{l+1}=\tfrac{H_l}{2},\;W_{l+1}=\tfrac{W_l}{2},\;D_{l+1}=\tfrac{D_l}{2},\;C_{l+1}=2\,C_l\
\end{equation}
\noindent This builds a low-resolution, high-semantic latent representation at the coarsest level $l = L-1$. 

The decoder mirrors this process in reverse and also has $L$ levels, with the same $n_l$ Transformer blocks per level as the encoder. Starting from the coarsest features, the feature maps are progressively upsampled and refined back to higher resolution. At each decoding level, it receives skip connections from the corresponding encoder level. Specifically, at decoder level $l$, the upsampled moving and fixed feature maps are merged with the skip connections $(\widetilde{\mathbf{F}}^l_m,\, \widetilde{\mathbf{F}}^l_f)$ from the encoder. Then, a linear fusion is applied to halve the channel dimensionality, and $n_l$ cross-attention Transformer blocks fuse and refine these features.
By the end of the decoder, feature maps with the identical patch resolution as level $0$ of the encoder are recovered. These two feature maps are finally concatenated along the channel dimension and passed to a U-Net style network for final flow prediction.
This dual-parallel stream architecture, with repeated cross-attention fusion, ensures that the output features encapsulate enriched information from both images across multiple domains and scales, facilitating precise deformation prediction for the U-Net architecture.

\begin{figure*}[ht]
    \centering
    \setcounter{subfigure}{0}
    \begin{overpic}[width=\textwidth]{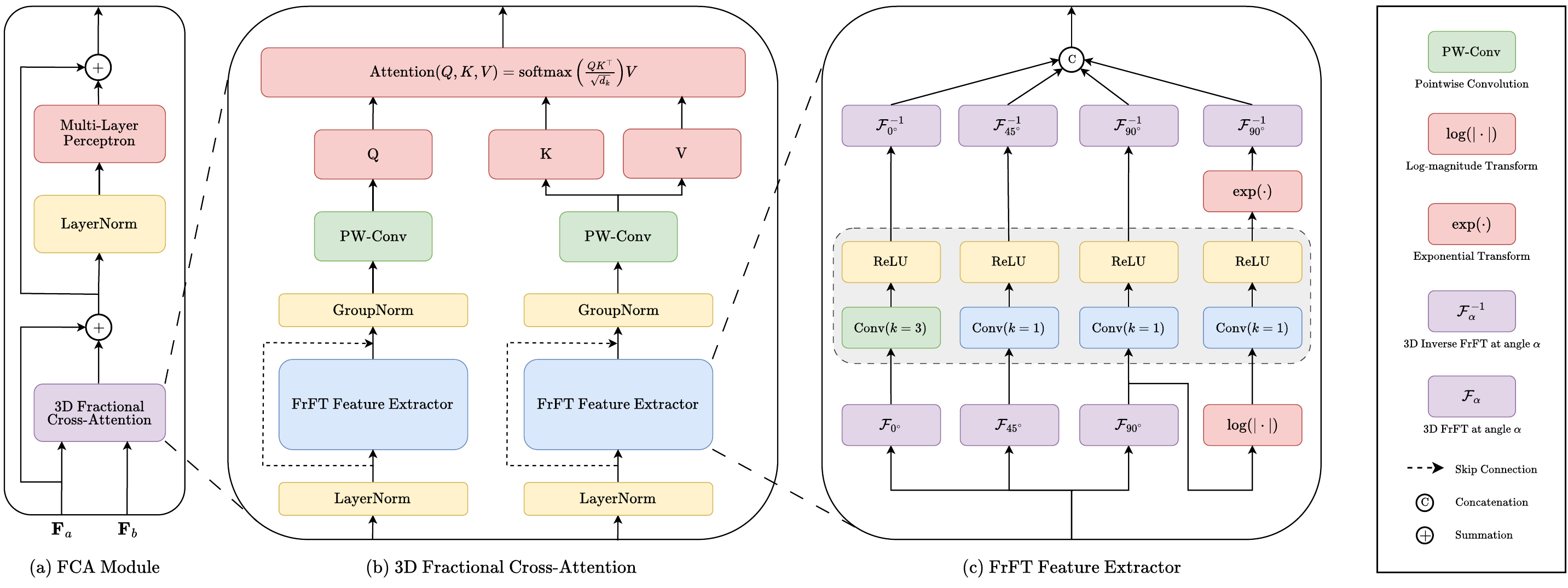}
        \put(12,75){\phantomsubcaption\label{fig:fca:a}}
        \put(50,75){\phantomsubcaption\label{fig:fca:b}}
        \put(88,75){\phantomsubcaption\label{fig:fca:c}}
    \end{overpic}
    \caption{
    (a) Overview of the proposed FCA module.
    (b) Internal structure of the Fractional Cross-Attention mechanism within the FCA module.
    (c) Architecture of the FrFT Feature Extractor, showing its four FrFT branches.
    }
    \label{fig:fca} 
\end{figure*}

\subsection{Fractional Fourier Transform}
\subsubsection{Theory and Fundamental Properties}
The FrFT generalizes the standard Fourier transform by introducing a fractional order parameter $p$, which corresponds to a rotation by angle $\alpha = p\pi/2$ in the combined spatial-frequency domain \citep{almeida1994fractional}. This makes the FrFT a powerful tool for observing different characteristics of a signal by varying its fractional degree $\alpha$. For a 1D signal $x(t)$, the $p$th-order FrFT $X_p(u)$ is defined by the integral transform:
\begin{equation}
\label{equ:1d-integral-frft}
X_{p}(u)
= \mathcal{F}_{p}\{x\}(u)
= \int_{-\infty}^{\infty} K_{p}(t,u)\,x(t)\,\mathrm{d}t
\end{equation}
\noindent where $K_p(t,u)$ is the fractional kernel. For $\alpha \neq n\pi$, this kernel is given by:
\begin{equation}
K_p(t,u)
\;=\;
A_\alpha\,
e^{\,j\!\bigl(\tfrac{t^2\cot\alpha}{2}
             -u\,t\,\csc\alpha
             +\tfrac{u^2\cot\alpha}{2}\bigr)}
\end{equation}
with $A_\alpha = \sqrt{\frac{1 - j\,\cot\alpha}{2\pi}}$ as a normalization factor. For the special cases where \(\alpha\) is an integer multiple of \(\pi\), the kernel reduces to a Dirac delta function. Specifically, \(K_p(t,u) = \delta(t - u)\) when \(\alpha = 2n\pi\), and \(K_p(t,u) = \delta(t + u)\) when \(\alpha = (2n \pm 1)\pi\). Also notably $\alpha = \pi/2$ ($p=1$) yields $\cot\alpha = 0$ and $\csc\alpha=1$, so the kernel reduces to $K_{p=1}(t,u) = e^{-j\,t\,u}$, which reproduces the standard Fourier transform formula.
The FrFT continuously interpolates between the spatial domain ($p=0$, no rotation) and the frequency domain ($p=1$, 90° rotation). In simple terms, a conventional image and its Fourier spectrum are just orthogonal extreme cases in the spatial-frequency domain. In contrast, the FrFT can represent the image at any intermediate rotation angle that simultaneously reflects both spatial and frequency characteristics of the image. This means applying a fractional transform by angle $\alpha$ is equivalent to taking the signal’s joint time-frequency energy distribution and rotating it by $\alpha$ in the plane \citep{ozaktas1996digital}. For $p=0.5$ (45° rotation), the representation is halfway between pure spatial and pure frequency view.

\subsubsection{Applications in Image Processing}
As previously noted, the fractional order $p$ controls the balance between spatial and frequency information in the transformed signal. This behavior is illustrated in Fig.~\ref{fig:frft-progress}, which exhibits how the FrFT progressively transforms input images as the fractional order increases. At $p = 0$, the transformation fully preserves the original spatial structure. At $p = 1$, the FrFT becomes the conventional Fourier transform, concentrating the image energy in the frequency domain and suppressing spatial details. At intermediate values, such as $p = 0.5$, the FrFT produces representations that retain both spatial and spectral features, as shown in the figure.
This property enables the capture of image characteristics ranging from coarse to fine across different fractional domains. In practice, extracting features at multiple fractional orders helps construct a rich, multi-scale feature set. It is well known that natural images, including medical images, are highly non-stationary, meaning that their local statistics, such as intensity distributions and textures, vary significantly across different regions. Here, the FrFT provides an advantage by capturing localized frequency components more effectively than the conventional Fourier transform. The FrFT in image processing has the ability to enhance feature extraction, especially for edges, oriented textures, and other direction-dependent features. Studies have shown that within the fractional Fourier domain, the phase component of an image carries rich texture and edge information \citep{zeng2015image}. This means that by tuning the FrFT order, different structural details of the image become pronounced. For instance, gradients and vessel-like structures in medical images can be better captured using FrFT-based filters or transforms.

Another practical advantage of using FrFT for image processing is its ability to suppress noise more effectively than a conventional Fourier approach in many situations. The key is that FrFT can target noise that is not uniform across the image (i.e. non-stationary noise). A classic Fourier transform is effective if noise
is concentrated at specific frequencies globally, but it struggles when noise characteristics vary over space. In contrast, the FrFT can adapt to such variations. Operating in a partial frequency domain, it can selectively isolate and suppress noise patterns that change with position \citep{subramaniam2010fractional, yang2012applications}. Figure~\ref{fig:spatial-spectral-space} illustrates this capability in the spatial–spectral representation of a non-stationary signal, where adjusting the fractional degree $\alpha$ enables effective separation of signal components from position-dependent noise.
Such capability is highly beneficial for tasks like deformable image registration, where noise can otherwise disrupt the alignment of images.
The above benefits make FrFT particularly useful in challenging image analysis domains like medical imaging. Medical images often contain structures with varying scales and orientations. For example, the heart muscle fibers have oriented textures, blood vessels form curvilinear patterns, and different tissues yield complex frequency content. FrFT-based methods have also shown promising results within the scope of image registration, as demonstrated in previous studies \citep{zhang2013medical}.

\begin{figure}[!t]
\centerline{\includegraphics[width=\columnwidth]{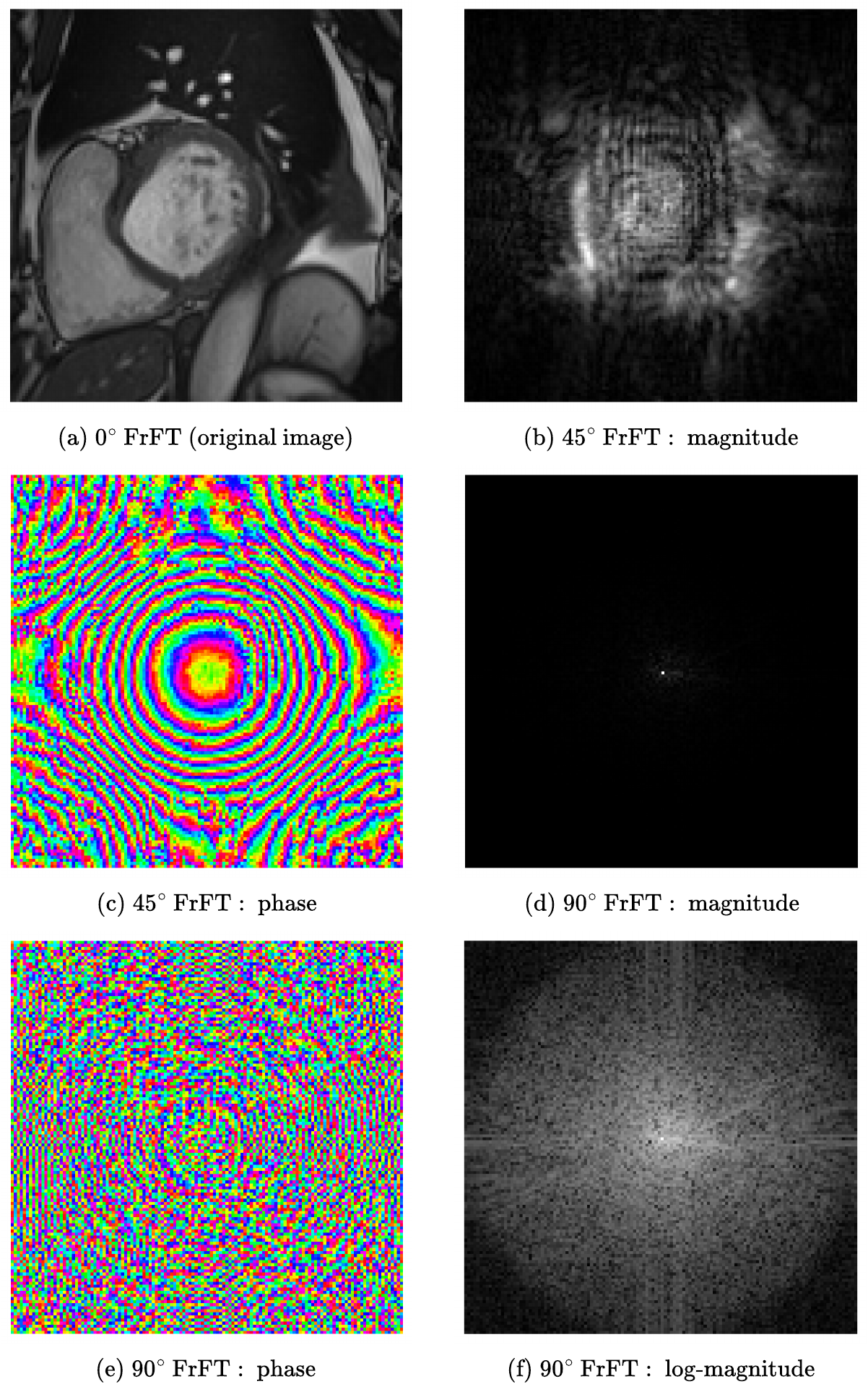}}
\caption{
FrFT outputs from the FCA feature extractor branches, shown before the convolution layers. The operations are applied to a 3D medical image. For illustration, only a single 2D slice of the volume is displayed.
(a) shows the result of \(\mathcal{F}_{0^\circ}\) operation.
(b)-(c) show the magnitude and phase of \(\mathcal{F}_{45^\circ}\) operation, respectively. 
(d)-(e) illustrate the magnitude and phase of \(\mathcal{F}_{90^\circ}\) operation, respectively.
(f) shows the logarithm of  \(\mathcal{F}_{90^\circ}\)'s magnitude.
}
\label{fig:frft-examples}
\end{figure}

\begin{table}[!t]
  \caption{Parameter counts and FLOPs of convolution operations in each branch of the FCA feature extractor module with channel coefficient $\alpha$, assuming an input tensor of size $C \times D \times H \times W$.}
  \label{tab:frft_branch_costs}
  \centering
  \scriptsize
  \renewcommand{\arraystretch}{1.2}
  \resizebox{\columnwidth}{!}{%
    \begin{tabular}{lcccc}
      \toprule
      Branch         & Kernel Size & Channels (In $\to$ Out)         & Parameters                     & FLOPs                                  \\
      \midrule
      FrFT0          & $3$           & $\alpha C \to \alpha C$    & $27\,\alpha^2 C^2$             & $27\,\alpha^2 C^2\,D\,H\,W$            \\
      FrFT45         & $1$           & $2\alpha C \to 2\alpha C$  & $4\,\alpha^2 C^2$              & $4\,\alpha^2 C^2\,D\,H\,W$             \\
      FrFT90         & $1$           & $2\alpha C \to 2\alpha C$  & $4\,\alpha^2 C^2$              & $4\,\alpha^2 C^2\,D\,H\,W$             \\
      Log-magnitude  & $1$           & $\alpha C \to \alpha C$    & $\alpha^2 C^2$                 & $\alpha^2 C^2\,D\,H\,W$                \\
      \midrule
      Total          & $-$           & $-$                           & $36\,\alpha^2 C^2$             & $36\,\alpha^2 C^2\,D\,H\,W$            \\
      \bottomrule
    \end{tabular}
  }
\end{table}

\subsubsection{Numerical Implementation and 3D Extension}
In our work, inspired by the implementation of the 1D FrFT by
\cite{pei1997improved} and the 2D FrFT developed by
\cite{yu2023deep}, we extend the FrFT to three-dimensional images. To implement the 3D FrFT, we take advantage of the fact that multidimensional transforms are separable \citep{pei1998two,sahin1995optical}. Thus, a 3D FrFT can be achieved by sequentially performing 1D fractional transforms along each axis (x, y, and z). We adopt the eigenvector‐decomposition approach for the 1D discrete FrFT.  A complete set of continuous FrFT eigenfunctions of order $p$ is given by the Hermite–Gaussian functions $\psi_n(x)$, which satisfy:
\begin{equation}
  \mathcal{F}_p\{\psi_n\}(x) \;=\; e^{-j\,p\,n\,\pi/2}\,\psi_n(x)
  \label{eq:cont-eig}
\end{equation}
and admit the closed‐form:
\begin{equation}
  \psi_n(x)
  = \frac{2^{1/4}}{\sqrt{2^n\,n!}}\,
    H_n\!\bigl(\sqrt{2\pi}\,x\bigr)\,
    \exp\!\bigl(-\pi\,x^2\bigr)
  \label{eq:hermite-gauss}
\end{equation}
where $H_n(\cdot)$ is the $n$th Hermite polynomial.  From Mehler’s formula, the continuous FrFT kernel can be written spectrally as:
\begin{equation}
  K_p(t,u)
  = \sum_{n=0}^{\infty}
    \psi_n(t)\,e^{-j\,p\,n\,\pi/2}\,\psi_n(u)
  \label{eq:cont-kernel}
\end{equation}

\noindent To obtain a discrete implementation, we truncate and sample Equation~\eqref{eq:cont-kernel} following
\cite{candan2000discrete}.  The resulting discrete FrFT of a length‑$N$ vector $x[n]$ is:
\begin{equation}
  X_{p}[m]
  = \mathcal{F}_{p}\{x\}[m]
  = \sum_{k=0}^{N-1} \mathcal{K}_{p}[m,k]\,x[k]
  \label{eq:disc-frft}
\end{equation}
\noindent with the spectral kernel:
\begin{equation}
  \mathcal{K}_{p}[m,n]
  = \sum_{k=0}^{N-1}
    u_{k}[m]\,e^{-j\,\tfrac{\pi}{2}p\,k}\,u_{k}[n]
  \label{eq:disc-kernel}
\end{equation}
where $u_k[n]$ are the discrete Hermite–Gaussian eigenvectors.

To compute the 3D FrFT of order $p$ on a volume $x[n_x,n_y,n_z]$, we apply three successive 1D FrFTs along the $x$-, $y$-, and $z$-axes:
\begin{equation}
\resizebox{.85\columnwidth}{!}{$
\begin{split}
X_p[m_x,m_y,m_z] = \mathcal{F}^{(z)}_{p}\!\left\{
    \mathcal{F}^{(y)}_{p}\!\left\{
        \mathcal{F}^{(x)}_{p}\{x\}
    \right\}
\right\}[m_x,m_y,m_z]
\end{split}
$}
\end{equation}
\noindent where 
$\mathcal{F}_p^{(x)}$, $\mathcal{F}_p^{(y)}$, and $\mathcal{F}_p^{(z)}$ denote applying the 1D FrFT of order $p$ along the $x$-, $y$-, and $z$-axes, respectively.
The inverse 3D FrFT is obtained by applying the inverse fractional transforms of order $-p$ along each axis:
\begin{equation}
\resizebox{.85\columnwidth}{!}{$
\begin{split}
x[n_x,n_y,n_z]
= \mathcal{F}^{(x)}_{-p}\!\left\{
    \mathcal{F}^{(y)}_{-p}\!\left\{
        \mathcal{F}^{(z)}_{-p}\{X_p\}
    \right\}
\right\}[n_x,n_y,n_z]
\end{split}
$}
\end{equation}
with $\mathcal{F}_{-p} = (\mathcal{F}_{p})^{-1}$.
This separable formulation offers superior computational efficiency compared to naive multi-dimensional FrFT implementations~\citep{yu2023deep}.

\begin{table*}[!t]
  \caption{Quantitative results on performance, model complexity, runtime, and memory on the ACDC dataset. Metrics are reported as mean~$\pm$~std for overall DSC, average DSC across structures, HD95, fraction of non-positive Jacobian voxels, and Jacobian std. FractMorph-Light and FractMorph are our proposed lightweight and primary models. \textbf{Bold} and \underline{underline} indicate best and second-best results, respectively.}
  \label{tab:comparison}
  \centering
  \scriptsize
  \renewcommand{\arraystretch}{1.2}
  \begin{tabular}{lcccccccc}
    \toprule
    Method
      & Overall DSC (\%)~$\uparrow$
      & Avg.\ DSC (\%)~$\uparrow$
      & HD95 (mm)~$\downarrow$
      & \%$|J_\phi|\le0$~$\downarrow$
      & STD($|J_\phi|$)~$\downarrow$
      & Parameters
      & Time (s)
      & Memory (MB) \\
    \midrule
    Initial
      & $80.77 \pm 7.05$
      & $58.41 \pm 11.96$
      & $2.73 \pm 1.09$
      & –
      & –
      & –
      & –
      & – \\
    \midrule
    ANTs (SyN)
      & $83.70 \pm 6.84$
      & $65.98 \pm 12.25$
      & $2.22 \pm 1.00$
      & $0.93 \pm 3.01$
      & $0.08 \pm 0.03$
      & –
      & $37.79$
      & $203$ \\
    Demons
      & $83.43 \pm 6.16$
      & $71.23 \pm 9.73$
      & $2.08 \pm 0.90$
      & $0.77 \pm 0.42$
      & $0.40 \pm 0.07$
      & –
      & $18.86$
      & $27$ \\
    VoxelMorph
      & $86.08 \pm 5.99$
      & $74.88 \pm 8.69$
      & $1.58 \pm 0.68$
      & $0.02 \pm 0.03$
      & $0.14 \pm 0.03$
      & $327{,}331$
      & $0.02$
      & $181$ \\
    Fourier‐Net
      & $83.09 \pm 6.01$
      & $60.68 \pm 12.09$
      & $2.36 \pm 0.97$
      & $\bm{0.00 \pm 0.00}$
      & $\bm{0.03 \pm 0.01}$
      & $879{,}677$
      & $0.02$
      & $38$ \\
    TransMorph
      & $82.87 \pm 7.22$
      & $68.11 \pm 11.70$
      & $2.05 \pm 0.88$
      & $0.06 \pm 0.01$
      & $0.15 \pm 0.05$
      & $46{,}689{,}459$
      & $0.30$
      & $491$ \\
    XMorpher
      & $84.96 \pm 4.04$
      & $70.09 \pm 9.53$
      & $1.82 \pm 0.77$
      & \underline{$0.02 \pm 0.04$}
      & \underline{$0.13 \pm 0.03$}
      & $15{,}093{,}891$
      & $0.46$
      & $207$ \\
    TransMatch
      & $86.19 \pm 4.76$
      & $47.25 \pm 8.54$
      & $3.21 \pm 1.59$
      & $0.03 \pm 0.04$
      & $0.14 \pm 0.03$
      & $112{,}262{,}611$
      & $0.38$
      & $1{,}008$ \\
    \midrule
    FractMorph‑Light
      & \underline{$86.32 \pm 4.86$}
      & \underline{$74.97 \pm 9.07$}
      & \underline{$1.57 \pm 0.82$}
      & $0.06 \pm 0.06$
      & $0.15 \pm 0.03$
      & $29{,}630{,}931$
      & $0.34$
      & $322$ \\
    FractMorph
      & $\bm{86.45 \pm 4.72}$
      & $\bm{75.15 \pm 8.95}$
      & $\bm{1.54 \pm 0.78}$
      & $0.05 \pm 0.04$
      & $0.15 \pm 0.03$
      & $63{,}910{,}483$
      & $0.36$
      & $461$ \\
    \bottomrule
  \end{tabular}
\end{table*}

\subsection{Fractional Cross-Attention Module}
At the core of each transformer block is our novel FCA module. The architecture of this module is illustrated in Fig.~\ref{fig:fca}. This module enhances the cross-attention mechanism by incorporating multi-domain feature extraction through the FrFT. Each block processes a pair of input feature tensors $\mathbf{F}^l_m$ and $\mathbf{F}^l_f$ of shape $\mathbb{R}^{H_l \times W_l \times D_l \times C_l}$. As implied by Fig.~\ref{fig:fca:b}, the FCA block consists of two main stages. The first stage is a multi-branch FrFT feature extractor that operates separately on each input. The second stage involves cross-attention, which matches the features between the two inputs. In the following sections, explanations of each stage are provided.

\subsubsection{Fractional Fourier Feature Extraction}
To enrich the features with both spatial and spectral representations, each normalized input is passed through four parallel branches, as illustrated in Fig.~\ref{fig:fca:c}. In each branch, a FrFT with a specific fractional order is applied to the input, followed by a convolution operation and a ReLU activation. The convolution kernel size is adapted to the domain. Specifically, in the spectral domain, a kernel size of $1$ is sufficient since, according to the convolution theorem \citep{bracewell1966fourier}:
\begin{equation}
\mathcal{F}\{f * g\} = \mathcal{F}\{f\} \cdot \mathcal{F}\{g\}
\end{equation}
\noindent where $\mathcal{F}\{\cdot\}$ denotes the Fourier transform, $*$ represents convolution in the spatial domain, and $\cdot$ denotes element-wise multiplication in the spectral domain. This property implies that each spectral coefficient already encodes global spatial information, making larger kernels unnecessary. In contrast, in the spatial domain, a kernel size of $3$ is used to effectively aggregate local neighborhood information. Finally, each branch passes through its corresponding inverse FrFT to transform the features back to the spatial domain, enabling the model to capture and enhance features in several FrFT domains.
According to the experiments by
\cite{yu2023deep}, fractional orders of $\alpha=0^\circ$, $\alpha=45^\circ$, and $\alpha=90^\circ$ are chosen for capturing local, semi-global, and global features, respectively. By selecting only these three fractional domains, our model facilitates effective feature enrichment while maintaining an appropriate level of computational complexity. Additionally, we adopt a log-magnitude branch that operates on the magnitude of the FrFT at $\alpha = 90^\circ$. As demonstrated by \citet{gonzalez2009digital}, the log-magnitude representation can reveal finer details within a transform, thereby enhancing the model’s ability to capture spectral information. Figure~\ref{fig:frft-examples} illustrates this effect, along with representative outputs from the different fractional Fourier transform branches. This transform is defined as $A = \log \big( 1 + | X_{90^\circ}(u) | \big)$.
After applying the convolution operation, the logarithm is inverted as $| X_{90^\circ}(u) | = \exp(A) - 1$.
After the four branch transformations, four distinct feature maps are obtained, each with the same shape as the input ($D_l \times H_l \times W_l \times C_l$). The outputs from the branches and the skip connection are concatenated along the channel dimension. Each branch is normalized, and after that, a pointwise convolution is applied to fuse them into a single unified representation. This fused feature map contains information from multiple complementary perspectives: spatial localized details, semi-global fractional domain, global spectral context, and detailed spectral magnitude.
By combining these, the network fully utilizes the advantages of both the spatial and frequency domains while maintaining feasible computational complexity, as shown in Table~\ref{tab:frft_branch_costs}. This table summarises the parameter counts and floating-point operations (FLOPs) for the convolution operations in each branch of the FCA feature extractor, assuming an input tensor of size $C \times D \times H \times W$ and channel coefficient $\alpha$. FLOPs measure the total number of arithmetic operations required to process data. Lower FLOPs generally indicate higher computational efficiency. The Parameters and FLOPs columns are computed from the kernel size and the number of input and output channels in each branch. 
We also propose two versions of our model: one in which each branch operates on the entire set of input channels, and another in which the input channels are split so that each branch processes only a portion of the channels, thereby reducing the model’s complexity. Formally, regarding Table~\ref{tab:frft_branch_costs}, for our main model we set the channel coefficient to $\alpha = 1$, which yields a total of $36\,C^2$ parameters and $36\,C^2 D H W$ convolutional FLOPs. For our lightweight model, we set $\alpha = \frac{1}{3}$, which reduces the overall branch parameters and convolutional FLOPs to $\frac{1}{9}$ of the main model.

\begin{table*}[!t]
  \caption{Quantitative results on performance, runtime, and memory on the LPBA40 dataset.}
  \label{tab:comparison-cerebral}
  \centering
  \scriptsize
  \renewcommand{\arraystretch}{1.2}
  \begin{tabular}{lccccccc}
    \toprule
    Method
      & Overall DSC (\%)~$\uparrow$
      & Avg.\ DSC (\%)~$\uparrow$
      & HD95 (mm)~$\downarrow$
      & \%$|J_\phi|\le0$~$\downarrow$
      & STD($|J_\phi|$)~$\downarrow$
      & Time (s)
      & Memory (MB) \\
    \midrule
    Initial
      & $82.06 \pm 2.82$
      & $54.79 \pm 4.32$
      & $6.28 \pm 1.74$
      & –
      & –
      & –
      & – \\
    \midrule
    ANTs (SyN)
      & $\bm{92.38 \pm 0.43}$
      & $\bm{71.83 \pm 0.98}$
      & $2.32 \pm 0.22$
      & $1.63 \pm 1.28$
      & $\bm{0.16 \pm 0.01}$
      & $303.87$
      & $214$ \\
    Demons
      & $86.23 \pm 0.19$
      & $61.68 \pm 3.51$
      & $2.70 \pm 1.07$
      & $\bm{0.04 \pm 0.01}$
      & $0.22 \pm 0.01$
      & $63.11$
      & $244$ \\
    VoxelMorph
      & $91.51 \pm 0.80$
      & \emph{$68.45 \pm 2.73$}
      & $1.20 \pm 0.35$
      & $0.51 \pm 0.05$
      & $0.29 \pm 0.02$
      & $0.26$
      & $3{,}608$ \\
    Fourier‐Net
      & $87.56 \pm 1.21$
      & $61.60 \pm 2.50$
      & $1.93 \pm 0.44$
      & \underline{$0.28 \pm 0.09$}
      & \underline{$0.20 \pm 0.02$}
      & $0.05$
      & $819$ \\
    TransMorph
      & \emph{$91.79 \pm 0.99$}
      & \underline{$68.68 \pm 2.75$}
      & $1.23 \pm 0.42$
      & $0.78 \pm 0.06$
      & $0.33 \pm 0.02$
      & $0.61$
      & $4{,}356$ \\
    XMorpher
      & $91.08 \pm 0.90$
      & $67.78 \pm 1.96$
      & $1.24 \pm 0.39$
      & $0.46 \pm 0.05$
      & $0.30 \pm 0.02$
      & $3.44$
      & $2{,}309$ \\
    TransMatch
      & $91.36 \pm 0.65$
      & $49.84 \pm 1.98$
      & $1.32 \pm 0.40$
      & $0.57 \pm 0.04$
      & $0.30 \pm 0.01$
      & $0.45$
      & $3{,}945$ \\
    \midrule
    FractMorph-Light
      & $91.76 \pm 0.69$
      & $68.26 \pm 2.81$
      & \underline{$1.12 \pm 0.26$}
      & $0.70 \pm 0.06$
      & $0.32 \pm 0.01$
      & $5.23$
      & $3{,}894$ \\
    FractMorph
      & \underline{$91.79 \pm 0.74$}
      & \emph{$68.52 \pm 2.58$}
      & $\bm{1.09 \pm 0.18}$
      & $0.53 \pm 0.03$
      & $0.29 \pm 0.01$
      & $5.26$
      & $4{,}037$ \\
    \bottomrule
  \end{tabular}
\end{table*}

\subsubsection{Cross‐Attention Mechanism}
After enriching each input feature map, cross-attention is performed to exchange information between the enriched moving feature map $O_{m}$ and the fixed feature map $O_{f}$. The spatial dimensions of each feature map are first flattened into a sequence of length \(N_l = D_l \cdot H_l \cdot W_l\). The query matrix \(Q_m \in \mathbb{R}^{N_l \times C}\) is computed from \(O_m\), while the key and value matrices \(K_f, V_f \in \mathbb{R}^{N_l \times C}\) are computed from \(O_f\), all using learned linear projections. The moving-to-fixed attention output is then computed as:
\begin{equation}
\text{Cross-Attention}(m \rightarrow f) = \text{softmax} \left( \frac{Q_m K_f^T}{\sqrt{d_k}} \right) V_f
\end{equation}
where $d_k = C/h$ is the dimension of the key vectors, with $h$ denoting the number of attention heads.
Similarly, in parallel, the reverse fixed-to-moving \(\text{Cross-Attention}(f \rightarrow m)\) is computed.
This cross-attention mechanism explicitly links features in $I_m$ and $I_f$ that are mutually relevant, in contrast to standard self-attention, which only considers intra-image relationships.
After the attended outputs are reshaped back to \(D_l \times H_l \times W_l \times C\), the original input corresponding to the query is added. This is followed by layer normalization, a feed-forward multi-layer perceptron, and a skip connection for each cross-attention, as illustrated in Fig.~\ref{fig:fca:a}.
This design encourages effective communication between the features of the two images at each level, while the fractional branches ensure that the attention mechanism has access to enriched feature representations spanning the spatial, mixed spatial-frequency, and frequency domains.

\subsection{Encoder-Decoder CNN for Deformation Field Generation}
The final stage of FractMorph is a lightweight U-Net style architecture \citep{ronneberger2015u} that generates the dense 3D deformation field $\phi$ from the fused Transformer features. After the dual Transformer decoder, feature maps for both the moving and fixed images are obtained. These feature maps are first concatenated and normalized. To restore the original spatial resolution, a reverse patch embedding is applied before the features are passed into the U-Net-style network. This network then processes these high-resolution features to generate the deformation field. Our U-Net architecture consists of three convolutional downsampling layers, a bottleneck, and three deconvolutional upsampling layers. It is a suitable choice because the Transformer generates multi-domain features that integrate local, semi-global, and global information. The U-Net architecture further refines these features to produce an accurate deformation field with well-preserved localized details.

\begin{figure*}[!t]
\centering
\includegraphics[width=\textwidth]{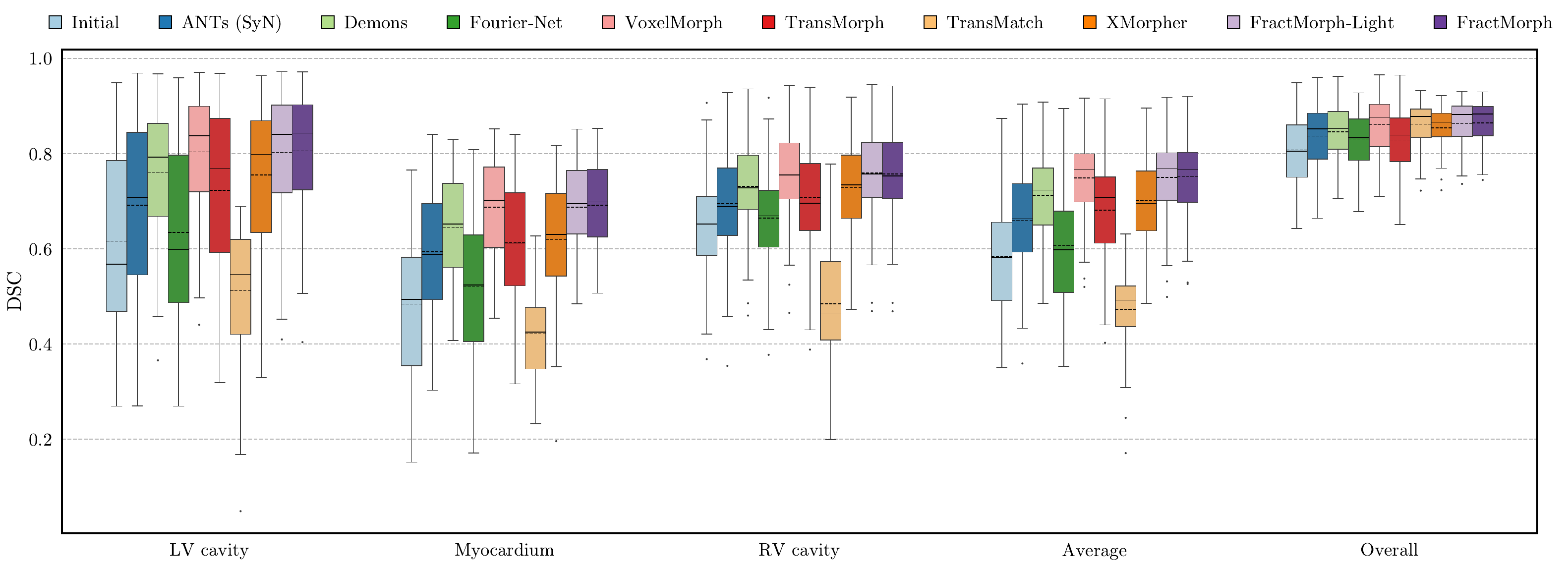}
\caption{Boxplots of Dice similarity coefficient for left ventricular cavity, myocardium, and right ventricular cavity segmentations, as well as the per–structure average and overall scores across nine registration methods on the ACDC dataset. FractMorph-Light and FractMorph are our proposed lightweight and primary models, respectively.}
\label{fig:boxplot_dsc}
\end{figure*}

\subsection{Spatial Transformation Function}
To calculate the similarity loss, the moving image \(I_m\) is warped using the predicted deformation field \(\phi\) to obtain the resulting warped image \(I_w\), which is then compared with the fixed image \(I_f\). To achieve this, we use a Spatial Transformation Function as proposed by
\cite{jaderberg2015spatial} and has also been widely adopted in DIR works such as \citep{balakrishnan2019voxelmorph, chen2022transmorph}.

For each voxel $p$, the deformation field $\phi$ maps it to a new location given by $p' = p + \phi(p)$. Since image intensities are defined only at integer voxel locations, trilinear interpolation is applied to estimate the intensity at \(p'\) using the values of its eight neighboring voxels:

\begin{equation}
    I_w(p) = I_m \circ \phi(p)
\end{equation}

\begin{equation}
   I_m \circ \phi(p) = \sum_{q \in \mathcal{N}(p')} I_m(q) 
\prod_{d \in \{x,y,z\}} (1 - |p'_d - q_d|) 
\end{equation}

\noindent where $\mathcal{N}(p')$ denotes the set of voxel neighbors of $p'$, and $d$ iterates over the three spatial dimensions. This differentiable interpolation allows gradients to flow through the warping operation, which enables the model to learn the optimal deformation field $\phi$ end-to-end.

\begin{figure*}[!t]
\centering
\centerline{\includegraphics[width=\textwidth]{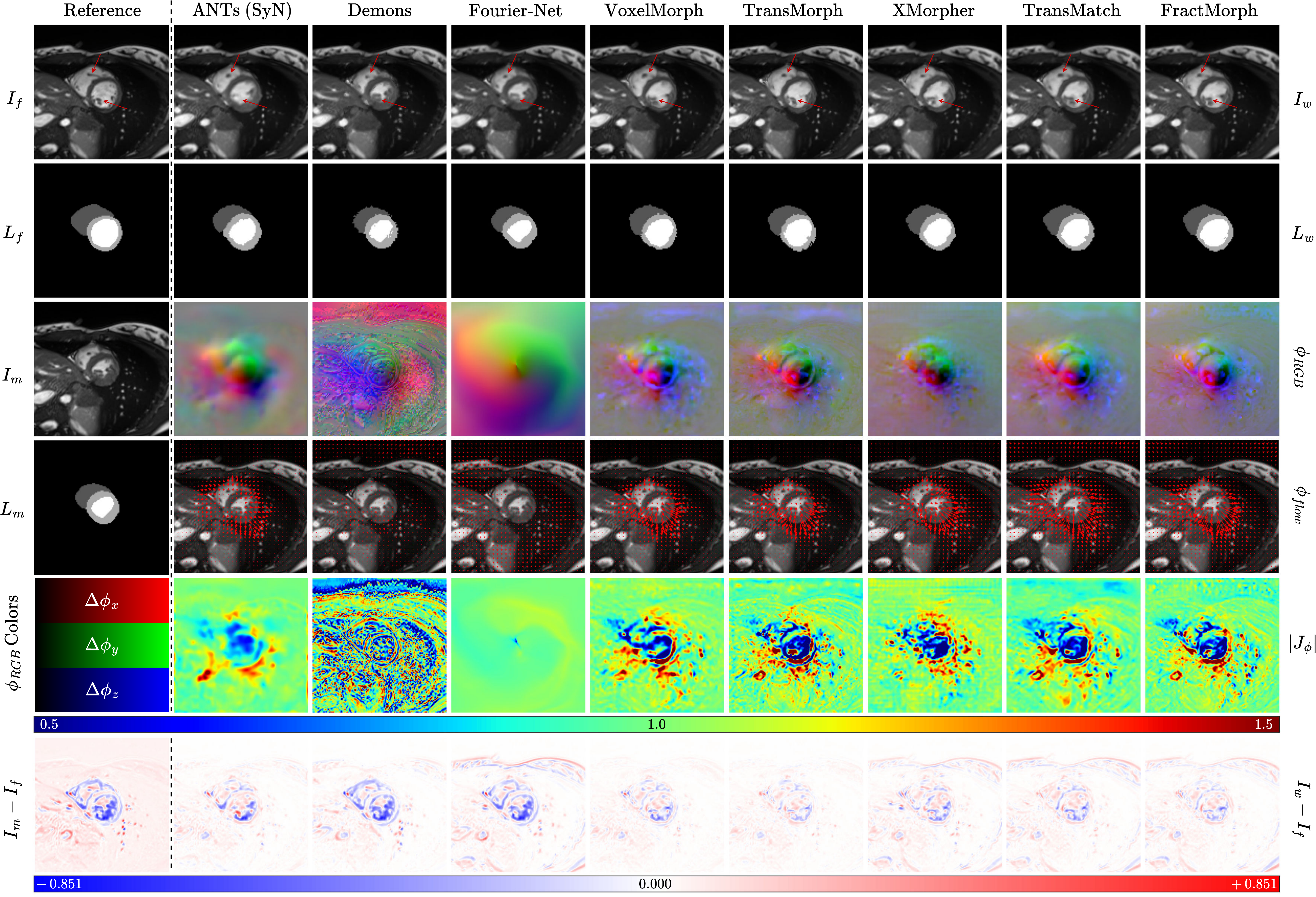}}
\caption{Qualitative comparison of registration methods applied to a fixed image ($I_f$) and a moving image ($I_m$), along with their corresponding anatomical segmentation labels ($L_f$ and $L_m$, respectively). From top to bottom: the warped image ($I_w = I_m \circ \phi$) and warped label ($L_w = L_m \circ \phi$), the generated deformation field (visualized in RGB), the 2D deformation flow along the $x$ and $y$ axes for the shown slice, the Jacobian determinant of the deformation field, and the difference between the warped and fixed images are shown. This example highlights the superiority of our model in capturing fine-grained local deformations, as indicated by the red arrows in the first row.}
\label{fig:sample-1}
\end{figure*}

\begin{figure*}[!t]
\centering
\centerline{\includegraphics[width=\textwidth]{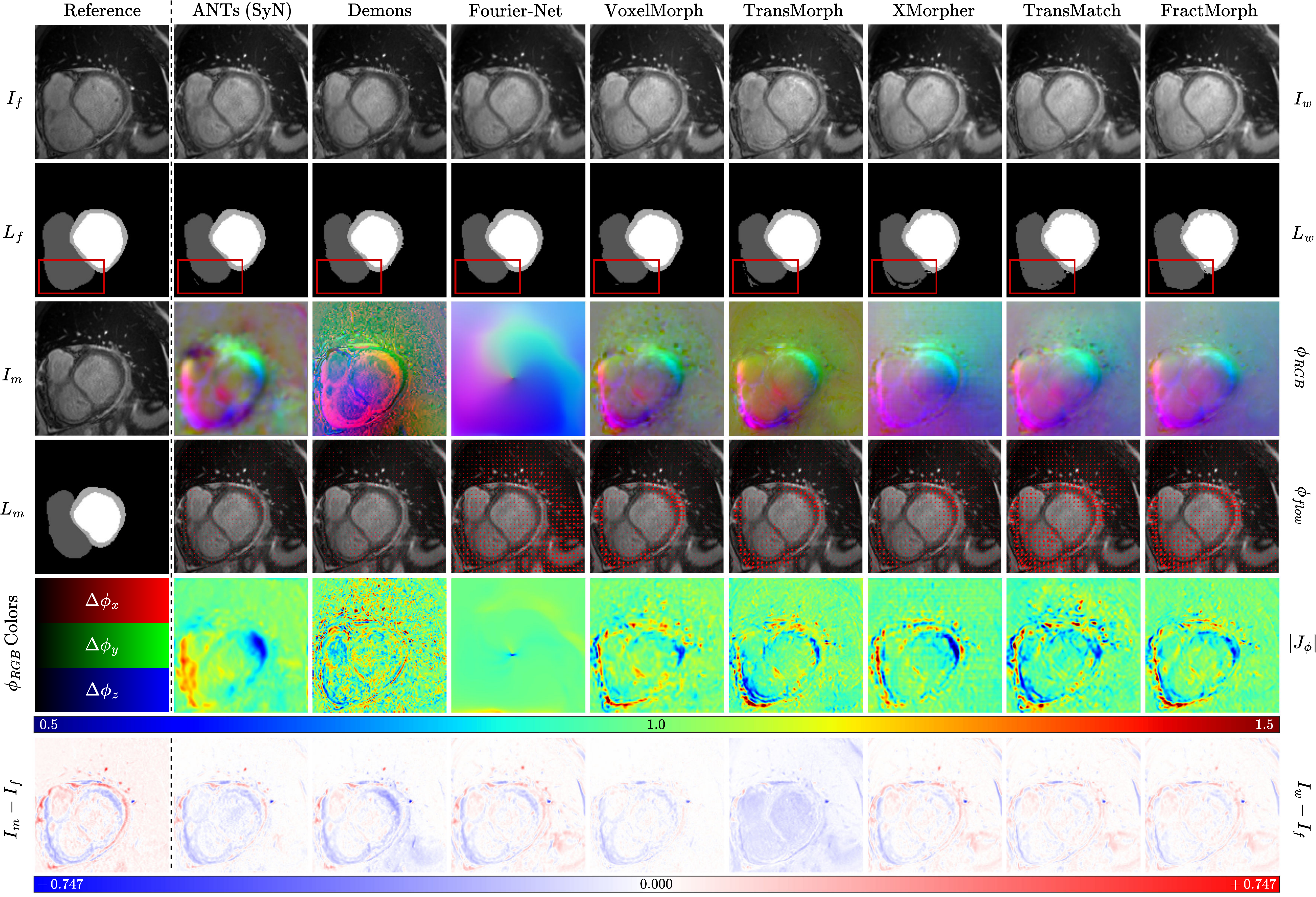}}
\caption{Qualitative comparison of registration methods on a case with semi-global deformations. This example illustrates the superior performance of our model in handling such deformations, as indicated by the red rectangles in the second row.}
\label{fig:sample-2}
\end{figure*}

\begin{figure*}[!t]
\centering
\centerline{\includegraphics[width=\textwidth]{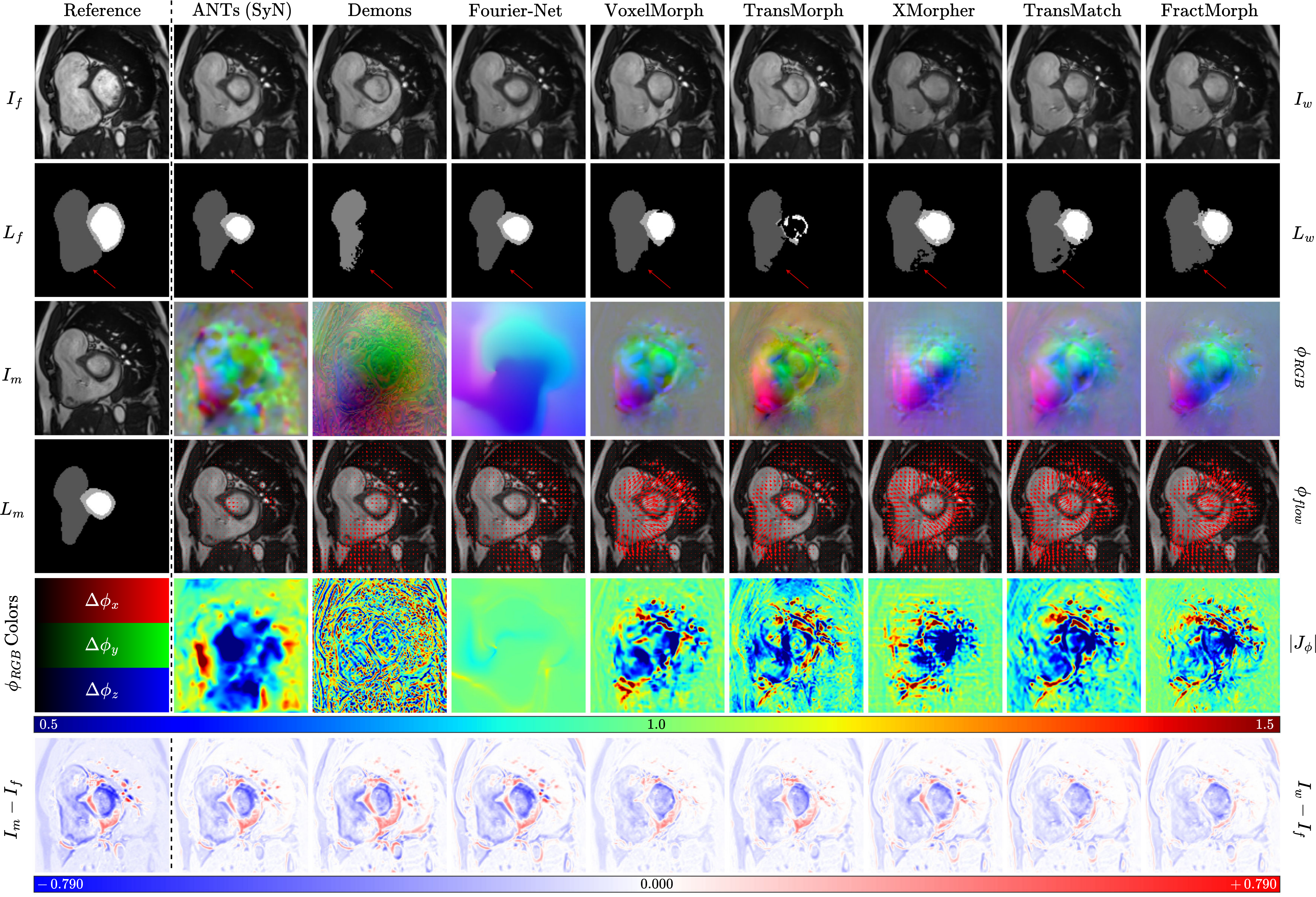}}
\caption{Qualitative comparison of registration methods on a case involving large-scale and global deformations. This example demonstrates the superior performance of our model in handling such deformations, which is clearly visible in the second row.}
\label{fig:sample-3}
\end{figure*}

\subsection{Loss Function}
\label{method:loss_function}
We train FractMorph in an unsupervised manner by combining an image similarity loss with a deformation regularization term. Adopting an unsupervised approach allows the model to be applicable to a wider range of datasets without requiring ground-truth deformation fields or anatomical segmentations. The total loss is defined as:
\begin{equation}
\mathcal{L}_{\text{total}}(I_f, I_m, \phi) = \mathcal{L}_{\text{similarity}}(I_f, I_m \circ \phi) + \lambda\, \mathcal{L}_{\text{smooth}}(\phi)
\end{equation}
where $(I_m \circ \phi)$ denotes the moving image $I_m$ warped by the deformation field $\phi$, producing the resulting image $I_w$. The function $\mathcal{L}_{\text{similarity}}(\cdot, \cdot)$ measures the similarity between the fixed image $I_f$ and the warped moving image $(I_m \circ \phi)$, while $\mathcal{L}_{\text{smooth}}(\cdot)$ encourages smoothness in the deformation field $\phi$. The parameter $\lambda$ controls the trade-off between similarity and smoothness.
In the following subsections, we explain each term in detail.

\subsubsection{Image Similarity Loss}

We use the local cross-correlation (CC) between the fixed image $I_f$ and the warped moving image $(I_m \circ \phi)$ as the similarity measure. The CC is robust to intensity variations across scans and datasets. Specifically, the local mean intensities $\hat{I}_f(p)$ and $\widehat{I_m \circ \phi}(p)$ are computed over a local $n^3$ neighborhood centered at voxel $p$:
\begin{equation}
\hat{I}_f(p) = \frac{1}{n^3} \sum_{p_i} I_f(p_i)
\end{equation}
\begin{equation}
\widehat{I_m \circ \phi}(p) = \frac{1}{n^3} \sum_{p_i} (I_m \circ \phi)(p_i)
\end{equation}

\noindent where $p_i$ iterates over the neighborhood. The local cross-correlation is then given by:
\begin{equation}
\resizebox{.85\columnwidth}{!}{$
\begin{split}
CC(I_f,I_m\circ\phi)=
\sum_{p\in\Omega}
\frac{\bigl(\sum_{p_i}[I_f(p_i)-\hat I_f(p)]\,[\,(I_m\circ\phi)(p_i)-\widehat{I_m\circ\phi}(p)\,]\bigr)^2}
     {\bigl(\sum_{p_i}[I_f(p_i)-\hat I_f(p)]^2\bigr)\,
      \bigl(\sum_{p_i}[(I_m\circ\phi)(p_i)-\widehat{I_m\circ\phi}(p)]^2\bigr)}
\end{split}
$}
\end{equation}
A higher CC indicates better alignment, so the similarity loss is defined as:
\begin{equation}
\mathcal{L}_{\text{similarity}}(I_f, I_m, \phi) = -CC(I_f, I_m \circ \phi)
\end{equation}

\subsubsection{Smoothing Regularization Term}

Minimizing only the similarity loss can lead to deformation fields that are non-smooth and unrealistic. To ensure that the estimated deformation field $\phi$ is spatially smooth, we add a diffusion regularizer on its spatial gradients. The smoothness loss is defined as:
\begin{equation}
\mathcal{L}_{\text{smooth}}(\phi) = \sum_{p \in \Omega} \| \nabla \phi(p) \|^2
\end{equation}
where $\nabla \phi(p)$ denotes the spatial gradient at voxel $p$:
\begin{equation}
\nabla \phi(p) = 
\left(
\frac{\partial \phi(p)}{\partial x}, 
\frac{\partial \phi(p)}{\partial y}, 
\frac{\partial \phi(p)}{\partial z}
\right)
\end{equation}
The partial derivatives are approximated by finite differences between neighboring voxels, for example:
\begin{equation}
\frac{\partial \phi(p)}{\partial x} \approx \phi(p_x + 1, p_y, p_z) - \phi(p_x, p_y, p_z)
\end{equation}
and similarly for 
$\frac{\partial \phi(p)}{\partial y}$ and $\frac{\partial \phi(p)}{\partial z}$

\section{Experiments and Results}
\label{sec:experiments}
\subsection{Datasets}
We evaluated our approach on the cardiac MRI ACDC \citep{8360453} and cerebral MRI LPBA40 \citep{shattuck2008construction} datasets. For ACDC, we performed intra-patient registration, while for LPBA40, we applied atlas-to-patient registration. 
The ACDC dataset contains cardiac cine MRI scans acquired at end-diastole (ED) and end-systole (ES). It includes 150 subjects, both healthy and with various cardiac pathologies. Expert annotations are provided for key cardiac structures: the left ventricular (LV) cavity, myocardium, and right ventricular (RV) cavity in both ED and ES phases. The dataset is divided into five groups of 30 cases each, categorized by physiological characteristics such as ventricular volumes, ejection fraction, local LV contraction, LV mass, and myocardial thickness. From each group, we selected 18 cases for training, 2 for validation, and 10 for testing, yielding 90 training cases, 10 validation cases, and 50 test cases. The test set corresponds to the official split proposed in \citep{8360453}, while training and validation cases were randomly sampled from the remainder. All scans were normalized to $[0,1]$ and resized to $16 \times 128 \times 128$, corresponding to a voxel size of $10 \times 1.8 \times 1.8$ mm. Ground-truth segmentations were used only for evaluation, as training was conducted in an unsupervised manner. 
The LPBA40 dataset comprises 40 T1-weighted brain MRI scans, each with 56 anatomical structures annotated by experts. All scans are skull-stripped, intensity-normalized, and resized to $160 \times 192 \times 160$. We split the dataset into 30 cases for training and 9 for testing, with one case designated as the atlas.

\subsection{Evaluation Metrics}
We evaluate the models using the Dice Similarity Coefficient (DSC) and the 95th-percentile Hausdorff Distance (HD95) to assess registration accuracy, as well as two additional metrics to measure the topological correctness of the deformation field. The standard deviation of each metric is also reported to demonstrate the stability of the models.
\subsubsection{Dice Similarity Coefficient}
The DSC is a region overlap measure used to quantify the similarity between two sets. Mathematically, for two binary sets $X$ and $Y$, the Dice coefficient is defined as:
\begin{equation}
DSC = \frac{2 |X \cap Y|}{|X| + |Y|}
\end{equation}
where $|X|$ and $|Y|$ denote the number of voxels in the annotated segmentations $X$ and $Y$, respectively. A higher DSC indicates greater overlap between the warped and fixed segmentations, and thus reflects more accurate registration.

\subsubsection{Hausdorff Distance}
The Hausdorff Distance (HD) is a boundary-based distance measure that evaluates the worst-case distance between two shapes. Given two sets of points, the HD is defined as the maximum of all nearest-neighbor distances from one set to the other. Formally, for two point sets $X$ and $Y$, the one-sided HD from $X$ to $Y$ is defined as:
\begin{equation}
hd(X, Y) = \max_{x \in X} \min_{y \in Y} d(x, y)
\end{equation}
where $d(x, y) = \| x - y \|_2$. To ensure symmetry, we use the bidirectional HD, as the one-sided HD is not commutative:
\begin{equation}
HD(X, Y) = \max \{ hd(X, Y), \, hd(Y, X) \}
\end{equation}
The HD95 is defined as the 95th percentile of all point-to-point distances between two surfaces. By excluding the most extreme 5\% of distances, HD95 reduces sensitivity to outliers. Therefore, we report HD95 as a robust boundary distance measure.

\subsubsection{Percentage of Non-positive Jacobian}
This metric evaluates the topological correctness of a deformation field by measuring the fraction of the spatial mapping that has a non-positive Jacobian determinant. The Jacobian matrix, defined as $J_{\phi}(p) = \nabla \phi(p)$, represents the local measurement of $\phi$ at voxel $p$. The Jacobian determinant describes the local volume change induced by the transformation at that point. To ensure a physically plausible deformation, it is required that \(|J_{\phi}(p)| > 0\) everywhere, thereby preventing local inversions or foldings. If $|J_{\phi}(p)| \le 0$ at any location, it indicates a fold at that voxel, meaning the mapping is not one-to-one in that neighborhood. Thus, a lower percentage reflects a deformation that better preserves topological consistency.

\subsubsection{Standard Deviation of Jacobian Values}
This metric measures the global variability of local volume changes caused by the deformation. It is computed as the standard deviation of the Jacobian determinant values across the entire image domain:
\begin{equation}
\sigma_{J} = \sqrt{ \frac{1}{N} \sum_{p \in \Omega} \Big( J_{\phi}(p) - \mu_{J} \Big)^2 }
\end{equation}
A lower $\sigma_{J}$ indicates a more uniformly smooth deformation, whereas a higher $\sigma_{J}$ suggests that the deformation contains regions with more extreme local volume changes, resulting in a less smooth deformation field.

\subsubsection{Model Size (Parameter Count)}
This metric measures the total number of trainable parameters in the model. Parameter count serves as an indicator of model capacity and the on‑disk storage required.
\subsubsection{Inference Time}
This metric measures the average time elapsed to complete a single forward pass on a pair of fixed and moving images.
\subsubsection{Memory Usage}
This metric measures the peak memory allocated during inference. We track RAM consumption for CPU‑based methods and VRAM usage for GPU‑based methods.

\subsection{Implementation Details}
We compare our framework with seven state-of-the-art registration methods, selected from traditional, non-transformer-based, and transformer-based approaches, to ensure a comprehensive evaluation. The learning-based models for the ACDC dataset were trained on an NVIDIA RTX 3080 GPU (12 GB VRAM), while those for the LPBA40 dataset, which required higher memory capacity, were trained on an RTX 4090 GPU (24 GB VRAM). To ensure fairness, evaluations of all methods were performed on the RTX 3080 GPU with identical software settings (CUDA 12.2, PyTorch 2.2.1).
Traditional methods, for which GPU-accelerated versions are not yet fully developed, were executed on an 11th Gen Intel Core i7-1165G7 CPU with 16 GB of 3200 MHz RAM.
For our framework, we use a patch size of $P_x = P_y = P_z = 4$ and a patch embedding dimension of $d=48$.
All methods were trained for 400 epochs using the Adam optimizer with a batch size of 1. We employed the loss function described in Section~\ref{method:loss_function}, using $\lambda = 1$ across all learning models to ensure a fair comparison. By setting \(\lambda = 1\), we equally weight the similarity term \(L_{\mathrm{similarity}}\) and the smoothness regularization \(L_{\mathrm{smooth}}\), balancing alignment accuracy against deformation smoothness and preventing either term’s scale from dominating. Keeping \(\lambda\) fixed also streamlines hyperparameter tuning and guarantees an unbiased, reproducible evaluation across all architectures.

\begin{table*}[!t]
  \caption{Ablation study of the FCA module and FrFT branches on the ACDC dataset, with mean~$\pm$~std reported.}
  \label{tab:ablation_frft}
  \centering
  \scriptsize
  \renewcommand{\arraystretch}{1.2}
  \begin{tabular}{cccc ccccc}
    \toprule
    \multicolumn{4}{c}{Branches}
      & \multicolumn{5}{c}{Results} \\
    \cmidrule(lr){1-4} \cmidrule(lr){5-9}
    $\mathrm{FrFT}_{0^\circ}$
      & $\mathrm{FrFT}_{45^\circ}$
      & $\mathrm{FrFT}_{90^\circ}$
      & $\log\bigl(|\mathrm{FrFT}_{90^\circ}|\bigr)$
      & Overall DSC (\%) $\uparrow$
      & Avg.\ DSC (\%) $\uparrow$
      & HD95 (mm) $\downarrow$
      & \%$|J_\phi|\le0$ $\downarrow$
      & STD($|J_\phi|$) $\downarrow$ \\
    \midrule
    $\times$ & $\checkmark$ & $\checkmark$ & $\checkmark$
      & $86.24 \pm 4.96$
      & \underline{$75.15 \pm 9.09$}
      & \underline{$1.56 \pm 0.83$}
      & $0.05 \pm 0.06$
      & $0.15 \pm 0.03$ \\

    $\checkmark$ & $\times$ & $\checkmark$ & $\checkmark$
      & $86.12 \pm 4.97$
      & $74.66 \pm 9.08$
      & $1.65 \pm 0.88$
      & $0.06 \pm 0.07$
      & $0.14 \pm 0.03$ \\

    $\checkmark$ & $\checkmark$ & $\times$ & $\checkmark$
      & $86.11 \pm 4.87$
      & $74.05 \pm 9.26$
      & $1.68 \pm 0.79$
      & $0.06 \pm 0.06$
      & $0.15 \pm 0.03$ \\

    $\checkmark$ & $\checkmark$ & $\times$ & $\times$
      & $86.07 \pm 5.10$
      & $74.66 \pm 9.04$
      & $1.60 \pm 0.81$
      & \underline{$0.05 \pm 0.05$}
      & $0.15 \pm 0.03$ \\

    $\checkmark$ & $\checkmark$ & $\checkmark$ & $\times$
      & \underline{$86.37 \pm 4.94$}
      & $74.89 \pm 9.35$
      & $1.57 \pm 0.83$
      & $0.06 \pm 0.06$
      & $\bm{0.14 \pm 0.03}$ \\

    $\checkmark$ & $\checkmark$ & $\checkmark$ & $\checkmark$
      & $\bm{86.45 \pm 4.72}$
      & $\bm{75.15 \pm 8.95}$
      & $\bm{1.54 \pm 0.78}$
      & $\bm{0.05 \pm 0.04}$
      & \underline{$0.15 \pm 0.03$} \\
    \bottomrule
  \end{tabular}
\end{table*}

\subsection{Baseline Methods}
In the following subsections, we briefly describe the seven state-of-the-art registration methods selected for comparison and the software packages used to run the traditional algorithms.

\subsubsection{ANTs (SyN)}
For the Symmetric Normalization (SyN) algorithm~\citep{avants2008symmetric} with mutual information as optimization metric, we used its Python wrapper package, ANTsPy\footnote{\url{https://github.com/ANTsX/ANTsPy}}. We applied the default recommended hyperparameters of the SyN registration function, but increased the number of iterations to (100, 80, 60) across the multi-resolution levels to achieve improved alignment accuracy.

\subsubsection{Demons}
For the Demons algorithm~\citep{thirion1998image}, we used the Python implementation provided by SimpleITK\footnote{\url{https://github.com/SimpleITK/SimpleITK}}. The Demons Registration Filter was configured with its default hyperparameters, except that we increased the Gaussian smoothing standard deviation of the displacement field to 5.0 and the number of iterations to 200, to improve convergence and reduce spurious deformations.

\subsubsection{VoxelMorph}
For VoxelMorph \citep{balakrishnan2019voxelmorph}, we used the official VoxelMorph-1 implementation\footnote{\url{https://github.com/voxelmorph/voxelmorph}} and trained it using the suggested learning rate of $1\mathrm{e}{-4}$.
\subsubsection{Fourier-Net}
For Fourier-Net \citep{jia2023fourier}, we used the official 3D Fourier-Net implementation\footnote{\url{https://github.com/xi-jia/Fourier-Net}} and trained it using the suggested learning rate of $1\mathrm{e}{-4}$.

\subsubsection{TransMorph}
For TransMorph~\citep{chen2022transmorph}, we employed the official implementation\footnote{\url{https://github.com/junyuchen245/TransMorph_Transformer_for_Medical_Image_Registration}}. We trained the model with the recommended learning rate of $1\mathrm{e}{-4}$.

\subsubsection{XMorpher}
For XMorpher \citep{shi2022xmorpher}, we used the official implementation\footnote{\url{https://github.com/Solemoon/XMorpher}} and trained it using the suggested learning rate of $1\mathrm{e}{-4}$.

\subsubsection{TransMatch}
For TransMatch~\citep{chen2023transmatch}, we employed the official implementation\footnote{\url{https://github.com/tzayuan/TransMatch_TMI}}. The model was trained using the recommended learning rate of $4e-4$.

\begin{figure*}[!t]
\centering
\includegraphics[width=\textwidth]{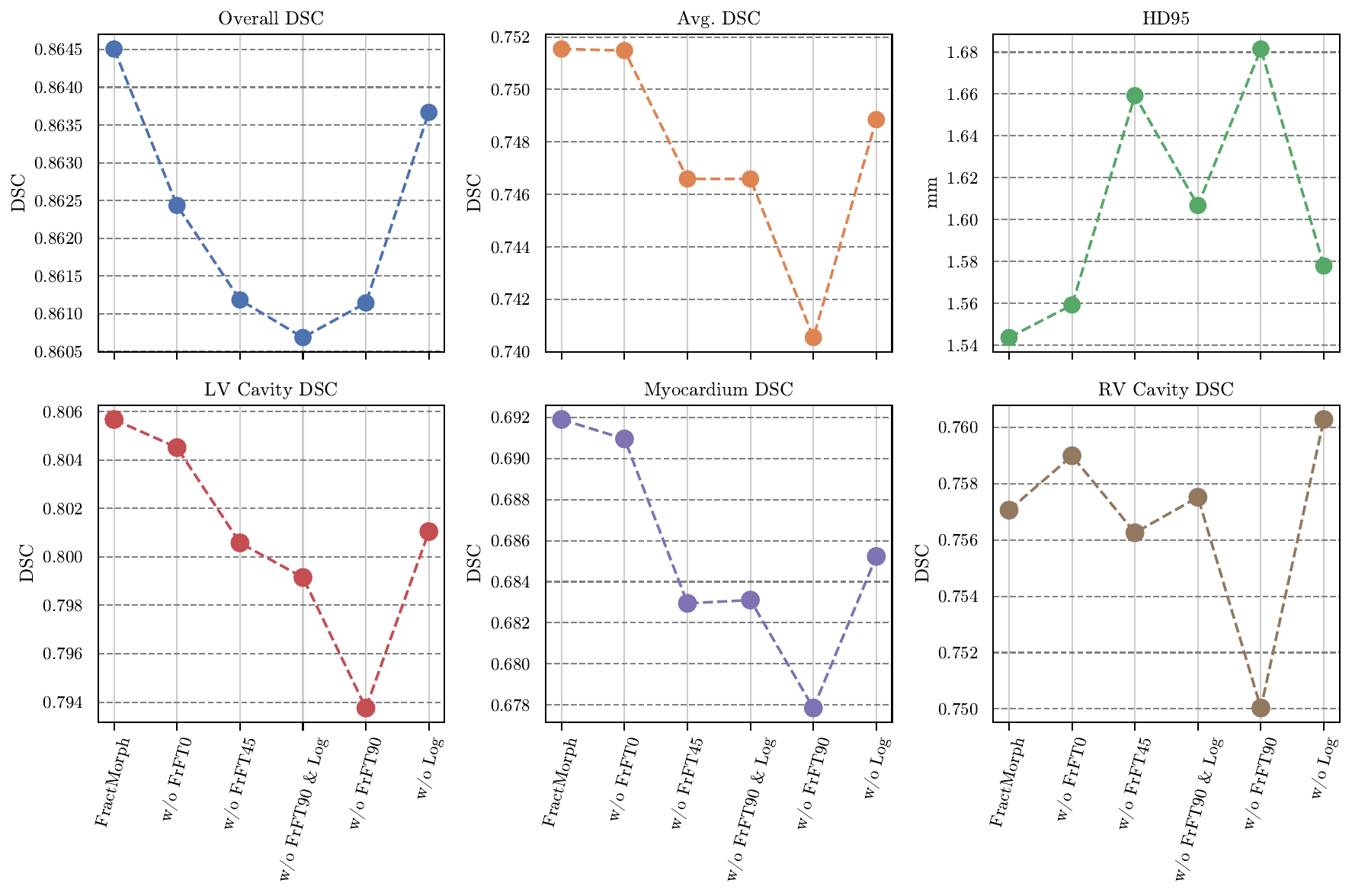}
\caption{Overall and per-structure registration accuracy in the ablation study showing the contribution of each FrFT branch on the ACDC dataset.}
\label{fig:frft-branches-contributions}
\end{figure*}

\subsection{Comparison to Baseline Methods}
\label{sec:comparison}

\subsubsection{Registration Accuracy}
\label{sec:accuracy}
As shown in Table~\ref{tab:comparison}, our proposed FractMorph framework achieved the highest performance across all registration accuracy metrics among the compared methods on the ACDC test set. FractMorph attained an overall DSC of 86.45\% and an average per-structure DSC of 75.15\%, which are marginally higher than the best baseline approaches. In particular, FractMorph improved the overlap of the cardiac structures compared to both traditional algorithms and prior learning-based models.
The lightweight variant of our model (FractMorph-Light) performed nearly on par with the full model, achieving an overall DSC of 86.32\% and an average DSC of 74.97\%. This indicates that even with reduced channels, our multi-domain attention approach retains a clear accuracy advantage over previous methods.
FractMorph and its lightweight variant also demonstrated superior boundary alignment accuracy, achieving the lowest HD95 values of 1.54 mm and 1.57 mm, respectively, thereby outperforming all baselines. In contrast, some competing methods showed larger HD95 values. For instance, TransMatch’s HD95 was 3.21 mm, indicating less consistent alignment at the boundaries despite its high overall DSC.

The superior precision of our method is evident across all anatomical structures. Fig.~\ref{fig:boxplot_dsc} presents boxplots of DSC for the LV cavity, myocardium, and RV cavity across all methods. Both FractMorph and its lightweight variant (FractMorph-Light) achieve higher mean DSC in every structure compared to the baselines. Moreover, FractMorph exhibits slightly higher means and lower variances than FractMorph-Light, indicating greater stability.
Our approach also preserves the topological correctness of deformations. The fraction of voxels with non-positive Jacobian determinants for FractMorph is 0.05\%, which is comparable to the best deep learning-based baselines and significantly lower than that of traditional iterative methods.
The Jacobian standard deviation for our method is 0.15, on par with other learning-based models. Notably, Fourier-Net achieved the lowest Jacobian variability (0.03) and zero foldings, but this came at the cost of substantially lower registration accuracy (83.09\% DSC).
In summary, FractMorph provides high alignment accuracy and physiologically plausible deformations, improving overlap across all structures while maintaining smooth and invertible transformations. Qualitative results further support these findings. 
Fig.~\ref{fig:sample-1} shows a representative case that reflects typical localized anatomical variability, demonstrating the overall registration quality achieved by our method. Fig.~\ref{fig:sample-2} includes a case with semi-global deformations, selected to illustrate the model’s ability to handle extended non-localized transformations that challenge baseline approaches. Fig.~\ref{fig:sample-3} highlights a case with large-scale and global deformations, representing scenarios where conventional models tend to break down.
Together, these examples demonstrate the robustness and generalization capability of FractMorph across diverse deformation patterns without the need for tuning to specific deformation scenarios.

The results above were obtained on an intra-patient registration task. In contrast, the cerebral dataset used for atlas-to-patient registration features more localized deformations compared to cardiac datasets, where detailed and large-scale deformations occur simultaneously. As shown in Table~\ref{tab:comparison-cerebral}, FractMorph achieved the best HD95 performance while maintaining substantially lower runtime compared to iterative models such as ATNs (SyN). Although SyN obtained the highest DSC scores, it required approximately 5 minutes per case, whereas our model achieved comparable Dice scores and superior HD95 in just 5 seconds. Notably, SyN also exhibited the highest number of non-positive Jacobians, indicating that its high DSC scores come at the cost of substantial folding and a lack of one-to-one transformations.

\subsubsection{Computational Efficiency}
\label{sec:efficiency}
The gains in accuracy with FractMorph come with a moderate increase in model complexity and runtime (see Table \ref{tab:comparison}). Our primary model contains approximately 63.9M parameters, which is larger than most CNN-based methods yet comparable to transformer-based models. The proposed FractMorph-Light variant significantly reduces the size to 29.7M parameters by using channel splitting (channel coefficient of $\alpha=1/3$) in the FrFT feature extractor. This nearly halves the model size while maintaining superior accuracy.
In terms of memory footprint, FractMorph requires approximately 461 MB during inference on a 3D volume pair. This is higher than simpler architectures such as Fourier-Net and VoxelMorph, but remains manageable on modern GPUs. Moreover, it is substantially lower than TransMatch, which requires around 1 GB, and also lower than larger models such as TransMorph. FractMorph-Light further lowers this consumption to 322 MB, which is also less than TransMorph’s and comparable to SyN, VoxelMorph, and XMorpher. Inference speed is also slightly affected by the richer multi-domain approach of our model. FractMorph registers a pair of 3D images in approximately 0.36 seconds. Although this is slightly slower than lightweight models such as VoxelMorph or Fourier-Net, it remains substantially faster than traditional iterative methods such as ANTs and Demons. Transformer-based baselines fall in between, with TransMatch requiring 0.38 seconds and XMorpher 0.46 seconds per case. Our lightweight variant, FractMorph-Light, provides a modest speed-up, completing registration in 0.34 seconds.
In the higher-resolution dataset of LPBA40, our models maintain superior runtime performance compared to iterative methods and also consume less memory than VoxelMorph, TransMorph, and TransMatch.

\begin{table*}[!t]
  \caption{Ablation study of the FCA module skip connection on the ACDC dataset, with mean~$\pm$~std reported.}
  \label{tab:ablation_skip}
  \centering
  \scriptsize
  \renewcommand{\arraystretch}{1.2}
  \begin{tabular}{c ccccc}
    \toprule
    Skip Connection
      & Overall DSC (\%)~$\uparrow$
      & Avg.\ DSC (\%)~$\uparrow$
      & HD95 (mm)~$\downarrow$
      & \%$|J_\phi|\le0$~$\downarrow$
      & STD($|J_\phi|$)~$\downarrow$ \\
    \midrule
    $\times$
      & $85.45 \pm 4.54$
      & $74.42 \pm 9.14$
      & $1.65 \pm 0.80$
      & $0.05 \pm 0.07$
      & $\bm{0.14 \pm 0.03}$ \\

    $\checkmark$
      & $\bm{86.45 \pm 4.72}$
      & $\bm{75.15 \pm 8.95}$
      & $\bm{1.54 \pm 0.78}$
      & $\bm{0.05 \pm 0.04}$
      & $0.15 \pm 0.03$ \\
    \bottomrule
  \end{tabular}
\end{table*}

\subsection{Ablation Study}
We conducted ablation studies to assess how removing key components affects the performance of our proposed model. First, we examine the contribution of each FrFT branch within the FCA module. Then, we demonstrate the necessity of the skip connection in the FCA module to achieve more effective registration.

\subsubsection{FrFT Branches}
\label{sec:ablation-frft}
To quantify each FrFT branch’s contribution, we disabled each branch in turn while keeping all other components unchanged (see Table \ref{tab:ablation_frft}). Removing any branch degrades accuracy, and the full model performs best overall. The branch $\mathrm{FrFT}{0^\circ}$ has the smallest effect with a 0.08\% drop in overall DSC and a 0.02 mm increase in HD95, while the branches $\mathrm{FrFT}{45^\circ}$ and $\mathrm{FrFT}{90^\circ}$ each cause over 0.33\% declines in overall DSC, over 0.11 mm increases in HD95, and average DSC losses of 0.49\% and 1.10\%, respectively. Fig.~\ref{fig:frft-branches-contributions} visualizes these trends across global metrics and per-structure DSCs, showing minimal impact from the $\mathrm{FrFT}{0^\circ}$ and log branches and maximal impact from $\mathrm{FrFT}{45^\circ}$ and $\mathrm{FrFT}{90^\circ}$. Notably, the log branch still improves every metric.
Crucially, all variants preserve diffeomorphic regularity, as the percentage of non-positive Jacobian determinants remains between 0.05\% and 0.06\%, and the standard deviation of $|J_\phi|$ ranges only from 0.14 to 0.15. The full model achieves the lowest folding rate (0.05\% ± 0.04) and a Jacobian-magnitude SD of 0.15 ± 0.03, demonstrating that adding branches boosts accuracy without compromising deformation smoothness.

\subsubsection{Skip Connection in FCA}
\label{sec:ablation-skip}
To assess the impact of the skip connection in our FrFT feature extractor, we compared models with and without it (see Table~\ref{tab:ablation_skip}). Adding the skip connection increases overall DSC by 1\%, average DSC by 0.73\%, and reduces HD95 by 0.11 mm. The rate of non-positive Jacobian determinants remains at 0.05\%, while the standard deviation of the Jacobian rises only marginally. These results show that skip connections significantly boost accuracy while maintaining diffeomorphic regularity.

\section{Discussion}
\label{sec:discussion}
FractMorph addresses a long-standing gap in deformable image registration: convolutional networks excel at modeling fine local details but lack global context, whereas standard transformers capture global structure at the expense of local precision and require large amounts of training data. To address these challenges, FractMorph brings together three key elements in a single end-to-end framework. First, we introduce a novel multi-domain FCA module that applies 3D FrFT branches at $0^\circ$, $45^\circ$, $90^\circ$, and a log-magnitude stream, enabling the network to extract local, semi-global, and global features in parallel. Second, we introduce a lightweight encoder–decoder CNN that transforms these transformer-enriched features into a high-resolution deformation field, preserving fine local details. Third, we adopt a dual-parallel transformer architecture to maintain continuous interaction between the fixed and moving image streams throughout feature extraction. Together, these components leverage their complementary strengths to achieve accurate and efficient deformable image registration.

Quantitatively (see Table~\ref{tab:comparison}), FractMorph achieves state-of-the-art accuracy on the ACDC cardiac cine-MRI benchmark, outperforming both classical algorithms and recent deep learning baselines on every registration metric. This gain does come with a modest runtime cost, as the FrFT operations introduce a small computational overhead that slightly increases inference time. However, our model still runs significantly faster than traditional optimization-based methods. Furthermore, the total parameter count (63.9 M for FractMorph, 29.6 M for FractMorph-Light) and memory consumption remain comparable to or even lower than those of leading transformer architectures. The memory footprint is also halved in the lightweight variant, making both models practical for deployment on modern and resource-constrained GPUs.
Qualitative examples in Figs.~\ref{fig:sample-1}, \ref{fig:sample-2}, and \ref{fig:sample-3} further illustrate FractMorph’s versatility. Whether the deformation is subtle and local or large and global, our single end-to-end network consistently produces smooth, topology-preserving warps without the need for scenario-specific tuning. This robustness stems directly from the multi-order FrFT branches, as each fractional angle captures features at a different spatial–frequency scale, as also explained in Section~\ref{sec:method}.

Our ablation study (see Table~\ref{tab:ablation_frft}) confirms the importance of each FrFT branch. Removing either the $45^\circ$ or the $90^\circ$ branch reduces overall DSC by up to $0.34\%$, average per-structure DSC by up to $1.10\%$, and increases HD95 by up to $0.14\,$mm. In Fig.~\ref{fig:frft-branches-contributions}, we visualize each branch’s contribution and see that all branches improve performance. However, the log-magnitude branch (without the $\alpha=90^\circ$ branch) can slightly degrade results, as the FrFT phase carries valuable information that the magnitude-only representation omits (see Section~\ref{sec:method}). Notably, adding the log-magnitude branch to the $\alpha=90^\circ$ branch yields better performance than the configuration with the $\alpha=90^\circ$ branch but without the log-magnitude branch, since the log transformation enables the model to better capture subtle details in the spectral magnitude domain. We also evaluated the residual skip connections in the FCA module and found that they further enhance feature enrichment and promote more stable gradient flow.
Another notable exception is the RV cavity, where omitting the $\alpha=0^\circ$ branch slightly increases DSC. In both the general population \citep{kawel2025society} and in our ACDC dataset, the RV cavity undergoes a large volume changes between end-diastole and end-systole and has a greater overall volume than the LV and myocardium on average. Omitting the $\alpha=0^\circ$ branch slightly reduces DSC for the LV and myocardium, whereas it slightly increases DSC for the RV cavity. We hypothesize that deprioritizing pure spatial features enables the network to focus more on semi-global and global patterns, improving alignment for larger structures like the RV cavity while compromising fine-detail registration in smaller structures.  
Apart from the cardiac and intra-patient datasets, we also evaluated our method on cerebral and atlas-to-patient datasets to demonstrate its versatility across different modalities and registration tasks. As shown in the Experiments section, our model achieves significantly better performance on cases involving local-to-global and multi-scale deformations. Nonetheless, it also performs well on other modalities with more localized deformations, making it a generally superior choice. Table~\ref{tab:comparison-cerebral} presents quantitative results supporting this claim.
We note that the LPBA40 dataset contains very few training cases, which poses a challenge for learning-based methods to accurately model deformations. Additionally, iterative methods like SyN are typically optimized for cerebral cases, potentially introducing a bias toward cerebral tissue, as our cross-modality experiments also suggest.

In summary, we demonstrated and analyzed the experiments and ablation results, highlighting the effectiveness and generality of our approach and FCA modules, as well as the necessity of each FrFT branch in building an end-to-end model capable of handling various types and scales of deformation.

\section{Conclusion}
\label{sec:conclusion}
In this work, we have presented FractMorph, the first 3D transformer-based architecture that integrates multi-domain FrFT branches into cross-attention to capture local, semi-global, and global deformations simultaneously. This framework addresses the challenge of modeling deformations at multiple scales within a fully end-to-end deformable image registration network. Our comprehensive evaluation on the cardiac intra-patient ACDC dataset shows state-of-the-art registration accuracy and anatomically plausible and smooth deformations, with feasible inference time and memory consumption. To further demonstrate the versatility of our method across different modalities and registration tasks, we also achieved high-performing results on the cerebral atlas-to-patient LPBA40 dataset. Ablation studies confirm the indispensable roles of each FrFT branch and skip connections. The lightweight FractMorph-Light variant highlights a practical trade-off between resource use and performance. Future work will focus on accelerating the FrFT operation to further improve runtime efficiency. In addition, the proposed FCA module holds promise for broader application across diverse tasks in both medical and non-medical imaging.

\section*{Declaration of Competing Interest}
The authors state that they have no financial or personal conflicts of interest that could have impacted the research presented in this manuscript.

\section*{Data Availability}
The ACDC dataset used in this study is publicly available at \url{https://www.creatis.insa-lyon.fr/Challenge/acdc/}, and the LPBA40 dataset can be accessed at \url{https://www.loni.usc.edu/research/atlas_downloads}.

\appendix

\printcredits

\bibliographystyle{cas-model2-names}

\printcredits

\bibliographystyle{cas-model2-names}


\end{document}